\documentclass[12pt,english]{article}
\usepackage[T1]{fontenc}
\usepackage[latin9]{inputenc}
\usepackage{float}
\usepackage{amsmath}
\usepackage{graphicx}
\usepackage{geometry}
\usepackage{esint}

\makeatletter
\newcommand{\lyxaddress}[1]{
	\par {\raggedright #1
	\vspace{1.4em}
	\noindent\par}
}

\@ifundefined{date}{}{\date{}}

\usepackage{xcolor}
\usepackage{array}
\usepackage{pifont}
\usepackage{wrapfig}
\usepackage{textcomp}
\usepackage{mathrsfs}
\usepackage{subcaption}
\usepackage{cite}

 \ifdefined\showcaptionsetup
 
\fi

\usepackage{babel}

\ifdefined\showcaptionsetup
 \PassOptionsToPackage{caption=false}{subfig}
\fi
\makeatother

\usepackage{babel}
\begin{document}
\title{Competitive binding of Activator-Repressor in Stochastic Gene Expression}
\author{Amit Kumar Das$^{1,2}$\thanks{mr.das201718@yahoo.com}, Debabrata
Biswas$^{1}$\thanks{debbisrs@gmail.com}}
\maketitle

\lyxaddress{$^{1}$Department of Physics, Bankura University, Bankura-722 155,
West Bengal, India.}

\lyxaddress{$^{2}$Digsui Sadhana Banga Vidyalaya, Digsui, Hooghly-712148, West
Bengal, India.}
\begin{abstract}
~In this paper, we explore the features of a genetic network where
the transcription factors (TFs), namely, activators and repressors
bind to the promoter in a \textit{competitive way}. We have developed
an analytical method to find the most probable set of parameter values
that are unavailable in experiments. We study the noisy behavior of
the circuit and compare the profile with the network where the activator
and the repressor bind the promoter \textit{non-competitively}. We
observe that the noise found in the super-Poissonian region of the
competitive genetic circuit is higher than the noise obtained in the
same from a non-competitive one. We further notice that, due to the
effect of transcriptional reinitiation in the presence of the activator
and repressor molecules, there exist some anomalous characteristic
features in the mean expressions and noise profiles. On top of that,
we find low noise in the transcriptional level and high noise in the
translational level in presence of reinitiation than in absence of
the same. In addition, we find out the method to reduce the noise
further below the Poissonian level in competitive circuit than the
non-competitive one with the help of some noise-reducing factors. 
\end{abstract}
\textit{Keywords}: Stochastic process, Gene Expression, Transcription
factor, The Fano factor, reinitiation, Poissonian level, Competitive
and non-competitive architecture

\section{Introduction}

\noindent Over the years, it has been experimentally verified and
accepted that the gene expression regulation is an inherent stochastic
process \cite{elowitz2002stochastic,ozbudak2002regulation,blake2003noise,raser2005noise,golding2005real,blake2006phenotypic,raj2006stochastic,suter2011mammalian,bartman2019transcriptional,thattai2004stochastic,biggar2001cell}.
In parallel with the experiments, many theoretical approaches, analytical
and numerical simulations, have been proposed in support \cite{paulson2005modelling,sanchez2008transcriptional,shahrezaei2008analytical,karmakar2004graded,karmakar2010conversion,kumar2014exact,bintu2005transcriptional,bintu2005transcriptionalapp,kuhlman2007combinatorial,vilar2003dna,vilar2005dna}.
Regulation of gene expression (GE) is essential for all organisms
to develop their ability to respond to the environmental changes.
It involves several complex stochastic mechanisms such as transcription,
pre-initiation, reinitiation, translation, and degradation, etc. \cite{kaern2005stochasticity,raj2008nature}.
Transcription is the initial step through which the biological information
is being transferred from the genome to the proteome \cite{alberts2002helper}.
The regulation of transcription \cite{bintu2005transcriptional,bintu2005transcriptionalapp}
occurs whenever some regulatory proteins called transcription factors
(TFs) interact with the promoter of a gene. Based on functionality,
TFs are classified as activators and repressors in both prokaryotes
and eukaryotes \cite{alberts2002helper,ptashne2005regulation,struhl1999fundamentally}.
Repressors inhibit the gene transcription by binding to specific DNA
sequences. Tryptophan and Lac are some widely known repressors for
prokaryotic systems. On the other hand, activators allow RNA-Polymerase
(RNAP) to bind to the promoter and initiate the transcription, resulting
in synthesizing mRNA. The molecular concentrations of these activators
and/or repressors can be varied with the help of some inducer molecules,
such as doxycycline (dox), galactose (GAL), tetracycline (Tc) and
anhydrotetracycline (aTc), etc. \cite{blake2003noise,thattai2004stochastic,biggar2001cell}.

An eukaryotic system is much more complex and possesses a compact
chromatin structure than prokaryotes. In the eukaryotic system, a
basic example of transcriptional regulation by TFs is a two-state
telegraphic model \cite{shahrezaei2008analytical,karmakar2004graded,kumar2014exact},
where a gene can be active (ON) or inactive (OFF) depending on whether
TFs are bound to the gene or not, respectively. An active gene transcribes
through the production of messenger-RNA (mRNA) in a pulsatile fashion
during short intense periods known as transcriptional bursts, followed
by a longer period of inactivity. This burst mechanism is the source
of cellular heterogeneity and stochastic noise \cite{raj2006stochastic,karmakar2004graded,karmakar2010conversion}.
Another crucial mechanism of transcription known as reinitiation \cite{blake2003noise,liu2014reinitiation}
is the cause of heterogeneity. This fact has been proven by experiments
\cite{blake2003noise,blake2006phenotypic,bartman2019transcriptional,shao2017paused,yudkovsky2000transcription}
as well as by recent theories \cite{cao2020stochastic,karmakar2020control,karmakar2021effect}.

\begin{figure}[H]
\centering{}\subfloat[]{\begin{centering}
\centering{}\includegraphics[width=6cm,totalheight=5cm,height=4cm]{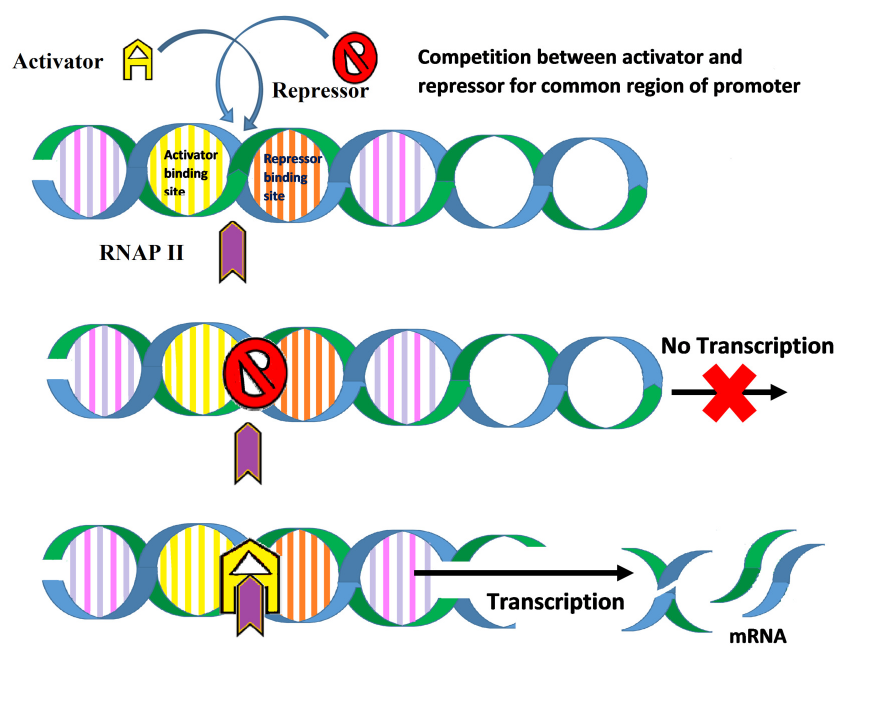}
\par\end{centering}
}\subfloat[]{\begin{centering}
\includegraphics[width=6cm,totalheight=5cm,height=4cm]{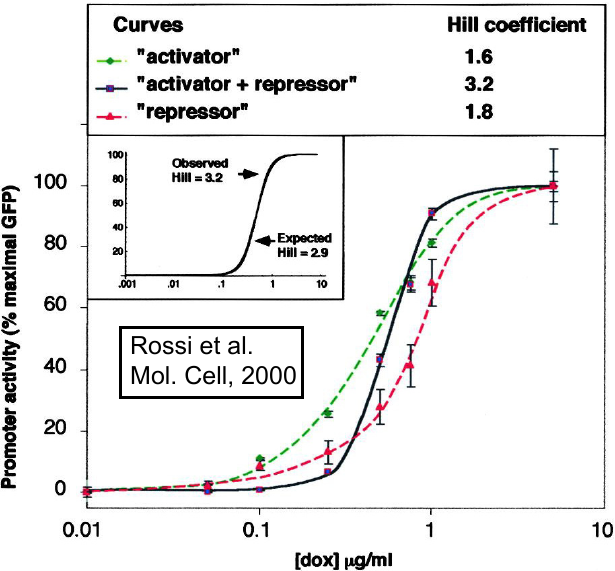} 
\par\end{centering}
}\caption{{\small (a) Competition between activator and repressor molecules
to bind to a common binding site of the promoter of gene {[}yellow
(red) region indicates activator (repressor) binding site{]} (b) Rossi
et al. \cite{rossi2000transcriptional} experimental plot: promoter
activity fitted with the Hill function.}}\label{fig:Competition}
\end{figure}

Most of the theories describe GE models using numerical simulations,
discrete stochastic models like the Gillespie algorithm \cite{gillespie1977exact}
and continuous stochastic models involving the chemical Langevin equation
(CLE) \cite{gardiner2009handbook}. The use of the Gillespie algorithm
and CLE are limited as they require details of the biochemical reaction
kinetics and rate parameters which are often unavailable in the performed
biological experiments. Due to these limitations, many biological
systems remain unmodeled or modeled without stochastic simulations.
One such experiment, carried out by Rossi et al. \cite{rossi2000transcriptional}
was not modeled. The authors used a dox-controlled synthetic transcription
unit driven by (i) activator only, (ii) repressor only, or (iii) both
in an overlapping promoter region. They proposed that the dose-response
curve follows the Hill function and the Hill coefficients from the
dose-response curves are observed as: 1.6 in the presence of the activator
only, 1.8 in the presence of the repressor only, and 3.2 when both
the activator and repressor operating together exclusively. Thus,
the Hill coefficient goes higher when activator and repressor compete.
Rossi et al. \cite{rossi2000transcriptional} suggested that either
addition or multiplication of the Hill coefficients 1.6 and 1.8 gives
3.4 and 2.88 ($\approx$ 2.9) respectively, which differs from what
they found to be 3.2 (Fig.~\ref{fig:Competition}b). Almost a decade
later, Yang and Ko \cite{yang2012stochastic} proposed a stochastic
Markov chain model (MCM) with a three-state activator-repressor system
to explain the deviation of observed Hill coefficients from the expected
values.

However, both these approaches were unable to explain the collective
(Hill coefficient) and stochastic behavior of the three-state competitive
activator-repressor system due to the lack of detailed kinetic rates
and parameter values. It is thus desirable to explain the model with
a proper theory supported by analytical and/or simulation methods.
With this as the motivation, accordingly, in this work, we develop
a theory that offers detailed chemical kinetics of the competitive
activator-repressor system and a most probable set of parameter values
that were due for decades. The availability of a suitable analytical
theory and the details of reaction kinetics could therefore be powerful
tools for future research and analysis of the genetic networks.

In this paper, we consider the competitive binding of TFs (activators
and repressors) to the promoter of the gene. There is evidence that
activators and repressors can regulate transcription by binding the
promoter of a gene mutually exclusively \cite{rossi2000transcriptional}.
It has been shown both experimentally \cite{biggar2001cell,yang2012stochastic,rossi2000transcriptional}
and analytically \cite{karmakar2010conversion,yang2012stochastic}
that the competition between activator and repressor molecules to
occupy the promoter region can generate a binary response while a
graded response is obtained when the molecules act independently separately.
However, there is no noise profile\footnote{Noise profile is represented by the Fano factor, often called as noise
strength. For more please see the glossary.} available for the transcriptional regulation by the activators and
repressors mutually exclusively in the above-mentioned works. Although
a few studies in Refs.~\cite{das2017effect,burger2012influence,soltani2015nonspecific},
have explored how the competitive binding of TFs in different scenarios
influences gene expression noise, nonetheless a general theoretical
framework of noisy behavior of a three-state activator-repressor system
still remains unexplored or may stick around in its infancy. Here,
we emphasize on studying the noisy behavior of the model and compare
its characteristics with a non-competitive activator-repressor model
\cite{das2022stochastic}. We will also check the behavior of the
architecture in the presence of different activator and repressor
molecules instead of doxycycline only.

The paper is organized in the following way. We provide a brief review
of the activator-repressor system in Sec.~\ref{sec:revactrep}. In
Sec.~\ref{sec:twostatAct} we consider the two-state model when only
the activator is present. The determination of the promoter activity
of the two-state system is provided by model analysis in Subsec.~\ref{subsec:modelanalysis}
and the corresponding parameter estimation is given in Subsec.~\ref{subsec:paraest_PA}.
The two-state model with only repressor is discussed in Sec.~\ref{sec:twostatrep}.
Next, we consider the competitive regulatory architecture in Sec.~\ref{sec:competitive},
wherein the parameter estimation for an activator-repressor system
is done in Subsec.~\ref{subsec:para_est_PAR}, and the stochastic
analysis is carried out in Subsec.~\ref{subsec:stochastic}. We compare
the competitive and the non-competitive architectures in Sec.~\ref{sec:comparison}.
The Subsec.~\ref{subsec:comparirole} finds out the role of reinitiation
under the action of $aTc$ and $GAL$ while Subsec.~\ref{subsec:noisered}
explores the effect of noise reducing factors. Finally, in Sec.~\ref{sec:conc}
we discuss the whole study and conclude with the future scopes.

\section{Activator-Repressor System: a brief overview}

\label{sec:revactrep}

The transcription factors such as activator and repressor molecules
can attach to the promoter of the gene and carry out their operation
in two ways: either competitively \cite{karmakar2010conversion} or
non-competitively \cite{das2022stochastic}.

In this paper, we have discussed the genetic circuits operated by
either activator only or repressor only or by both mutually exclusively.
Generally, the activator molecules get attached to the promoter of
a gene at a specified region known as the activator-binding site (yellow
bars in Fig.~\ref{fig:Competition}a). They subsequently make the
gene ready (active/ON) for transcription by allowing RNAP-II to bind
with them. On the other hand, when the repressor molecules occupy
their binding site (indicated by red bars in Fig.~\ref{fig:Competition}a),
they immediately block RNAP-II from sitting on the promoter. As a
result, the gene goes into the inactive/OFF state, \textit{i.e.},
inhibits transcription.

In recent work \cite{das2022stochastic}, we have rigorously studied
the architecture and the noise profile of the non-competitive binding
by activator and repressor molecules. The dose (dox)-response (green
fluorescent protein/GFP) of an competitive activator-repressor system
was experimentally shown in \cite{rossi2000transcriptional} and fitted
the parameters with the Hill function. Later, Yang and Ko \cite{yang2012stochastic}
using stochastic simulation tried to explain the mismatch of the Hill
coefficient with the observed experimental data. Furthermore, Karmakar
\cite{karmakar2010conversion} calculates the Probability Distribution
Function (PDF) for protein level of a three-state stochastic activator-repressor
network and explains the transformation of graded to the binary response
of PDF as observed in the experiment of Rossi et al. \cite{rossi2000transcriptional}.
However, none but \cite{karmakar2010conversion}, provide the details
of the kinetics and parameters of the system. Clearly, there is a
lack of exact parameter values for this network.

To find the parameter values we use a two-state and a three-state
genetic network operated by doxycycline as activator and/or repressor.
It must be stressed that, the aforementioned works have used networks
that show protein synthesis directly from genes and skipped the intermediate
transcription stages (mRNA synthesis). Here, we rigorously mention
all the possible transitions and parameters involved in the system.
However, to fit the parameters with the experimental data, we use
the mRNA dynamics instead of protein (one can consider the transcriptional
and translational stages have merged to a single stage since it is
well known that dynamics followed by mRNA reflect exactly that of
the protein multiplied by a scale factor to the magnitudes \cite{karmakar2020control,karmakar2021effect}),
as it reduces the number of parameters and their mathematical complexities.
Therefore we must multiply some scale factor to the estimated parameter
values to fit them with the experimental data (Fig.~\ref{fig:AR Curve-fitting_all}b).

In this study, we observe some interesting dynamical behavior (stochasticity
and noise) when both activator and repressor compete mutually for
a common binding region of the promoter. In addition, we compare the
characteristic differences between this competitive model and a non-competitive
model \cite{das2022stochastic}.

\section{Two-state model when only activator is present}

\label{sec:twostatAct}

\begin{figure}[H]
\centering{}\subfloat[]{\begin{centering}
\centering{}\includegraphics[width=7cm,totalheight=4cm,height=4cm]{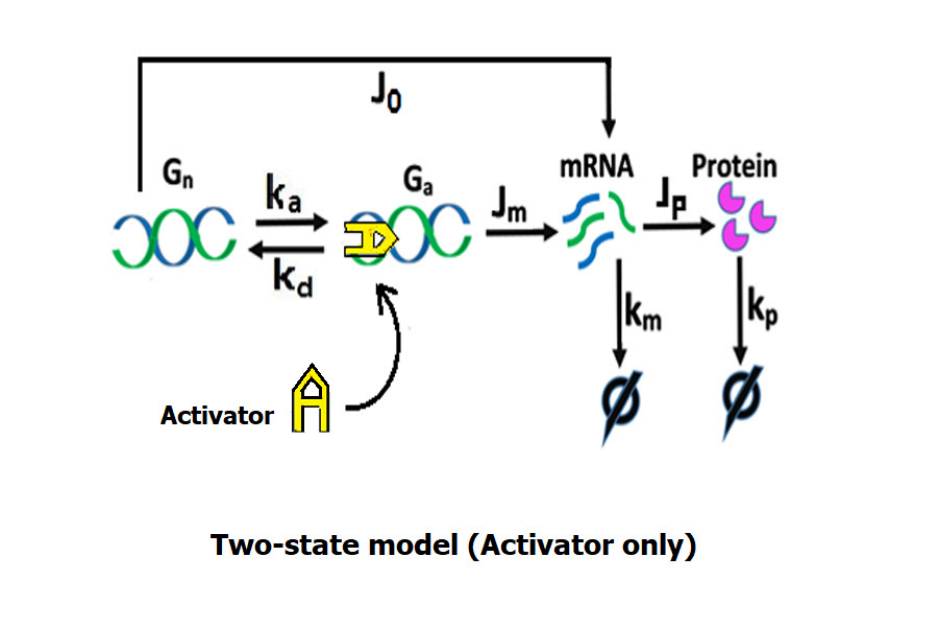}
\par\end{centering}
}\subfloat[]{\begin{centering}
\centering{}\includegraphics[width=6cm,totalheight=4cm,height=4cm]{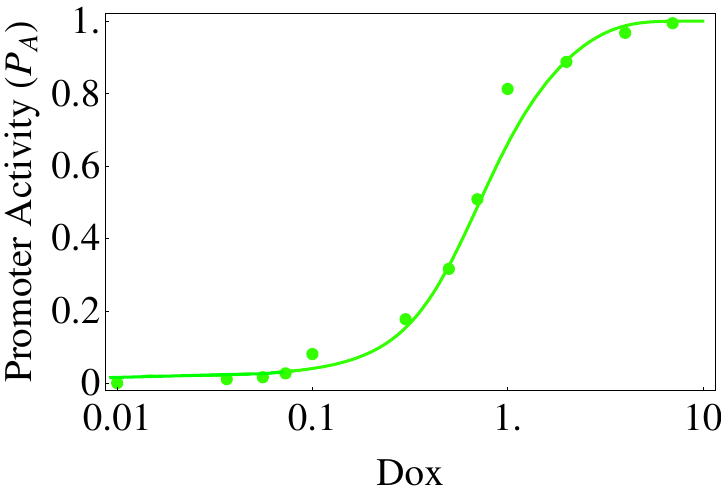}
\par\end{centering}
}\caption{{\small (a) Two-state system where only activator is operating (b)
Dose-response curve (solid curve) fitted with experimental data (solid
circles) in the presence of activator only (repressor absent) in the
promoter of gene.}}\label{fig:2-Two-state-activator}
\end{figure}

We start with a two-state model \cite{shahrezaei2008analytical,karmakar2004graded,kumar2014exact}
(Fig.~\ref{fig:2-Two-state-activator}a) where mRNAs may be produced
spontaneously from a normal state of gene $G_{n}$ with a very low
leaky basal rate $J_{0}$. The gene can be activated by some inducer
(activator), like, doxycycline (Dox) or galactose (GAL) or by some
protein produced by the gene itself (positive feedback). The mRNAs
are being produced from the active state $G_{a}$ (ON) at a rate $J_{m}=J_{1}.$
This mechanism of producing mRNA is called transcription. Protein
synthesis starts from the new born mRNAs via rate constant $J_{p}$
by the process, commonly known as translation. Both mRNAs and proteins
degrade by rates $k_{m}$ and $k_{p}$, respectively.~

\subsection{Model analysis: determination of promoter activity}

~\label{subsec:modelanalysis} The time evolution of mRNA concentration
can be expressed deterministically as a function $f(z,G)$ of genetic
states ($G$) as 
\begin{equation}
\frac{dz}{dt}=\frac{J_{i}}{z_{max}}G-k_{m}z=f(z,G),\label{eq:1 time evolution of mRNA}
\end{equation}
where $i=0$ (OFF) or = 1 (ON) such that $G=0$ (gene is inactive)
and G = 1 (gene is active), respectively. The variable $z$ denotes
the concentration of mRNA at time $t$ normalized by the maximum possible
mRNA concentration ($z_{max})$. Here we assume that only stochasticity
is involved in the transition of promoter between ON and OFF states.

The probability distribution function (PDF) can be written in the
form of the Fokker-Planck equation as 
\begin{equation}
\frac{\partial P_{i}(z,t)}{\partial t}=-\frac{\partial}{\partial z}\left[f(z,G)P_{i}(z,t)\right]+\sum_{j\neq l}{\displaystyle k_{jl}P_{j}(z,t)-k_{lj}P_{l}(z,t)}.\label{eq:2 Fokker-Planck eq}
\end{equation}
The first term in RHS is the transport term, second one is the gain
term and the last one implies the loss term. Using Eqs.~(\ref{eq:1 time evolution of mRNA})
and ~(\ref{eq:2 Fokker-Planck eq}) the Champman-Kolmogorov equation
for the reaction scheme of Fig.~\ref{fig:2-Two-state-activator}
can be written as 
\begin{equation}
\frac{\partial P_{0}(z,t)}{\partial t}=-\frac{\partial}{\partial z}\left[\left(\frac{J_{0}}{z_{max}}-k_{m}z\right)P_{0}(z,t)\right]+k_{d}P_{1}(z,t)-k_{a}P_{0}(z,t),~~~~\label{eq:pdf of 0 state}
\end{equation}
\begin{equation}
\frac{\partial P_{1}(z,t)}{\partial t}=-\frac{\partial}{\partial z}\left[\left(\frac{J_{1}}{z_{max}}-k_{m}z\right)P_{1}(z,t)\right]+k_{a}P_{0}(z,t)-k_{d}P_{1}(z,t),~~~~\label{eq:pdf of 1 state}
\end{equation}
we have the continuity equation 
\begin{equation}
\frac{\partial P}{\partial t}=-\nabla\cdot\mathcal{J}.\label{eq:continuity eq}
\end{equation}

In the next step, using no flux boundary condition we get the probability
current density as 
\[
\mathcal{J}=\left[\frac{J_{i}}{z_{max}}-k_{m}z\right]P_{i}=0.
\]
Also the steady state condition is given by 
\[
P(z)=P_{1}(z)+P_{0}(z).
\]
We have calculated the solution of the coupled Eqs.~(\ref{eq:pdf of 0 state})
and ~(\ref{eq:pdf of 1 state}) as 
\begin{equation}
P(z)=C_{1}\left(z-\frac{J_{0}}{k_{m}z_{max}}\right)^{\frac{k_{a}}{k_{m}}-1}\left(\frac{J_{1}}{k_{m}z_{max}}-z\right)^{\frac{k_{d}}{k_{m}}-1},\label{eq:probability}
\end{equation}
where, $z_{max}=\frac{J_{1}k_{a}+J_{0}k_{d}}{(k_{a}+k_{d})k_{m}}=$
steady state mean mRNA.

The normalization constant $C_{1}$ can be obtained by integrating
$P(z)$ from $z_{min}=\frac{J_{0}}{k_{m}}$ to $z_{max}$, \textit{i.e.},
$\int_{z_{min}=\frac{J_{0}}{k_{m}}}^{z_{max}}P(z)dz$.

Now the steady state probability of finding a cell with $z>Z_{th}$,
where $Z_{th}$ is a threshold value, is given by 
\begin{equation}
P(z>Z_{th})=1-\frac{\intop_{z_{min}}^{Z_{th}}P(z)\ dz}{\intop_{0}^{1}P(z)\ dz}=P_{A}\thinspace(\mbox{say}),\label{eq:promoter activity definition}
\end{equation}
\begin{equation}
P_{A}=1-\frac{\left(Z_{th}-J_{0}\right){}^{k_{\text{a}}}\left(J_{1}-Z_{\text{th}}\right){}^{k_{\text{d}}}{}_{2}F_{1}\left(Q_{1}\right)-\left(z_{min}-J_{0}\right){}^{k_{\text{a}}}\left(J_{1}-z_{min}\right){}^{k_{\text{d}}}{}_{2}F_{1}\left(Q_{2}\right)}{\left(1-J_{0}\right){}^{k_{\text{a }}}\left(J_{1}-1\right){}^{k_{\text{d}}}{}_{2}F_{1}\left(Q_{3}\right)-\left(-J_{0}\right){}^{k_{\text{a}}}J_{1}^{k_{\text{d}}}{}_{2}F_{1}\left(Q_{4}\right)},\label{eq:promoter activity Activator}
\end{equation}
all the parameters in Eq.~(\ref{eq:promoter activity Activator})
are normalized by $k_{m}$ and here $_{2}F_{1}\left(Q_{j}\right)$
is the Hypergeometric function given as 
\[
_{2}F_{1}\left(Q_{1}\right)={}_{2}F_{1}\left(1,k_{\text{a}}+k_{\text{d}};k_{\text{d}}+1;\frac{Z_{\text{th}}-J_{1}}{J_{0}-J_{1}}\right),
\]
\[
_{2}F_{1}\left(Q_{2}\right)={}_{2}F_{1}\left(1,k_{\text{a}}+k_{\text{d}};k_{\text{d}}+1;\frac{z_{\text{min}}-J_{1}}{J_{0}-J_{1}}\right),
\]
\[
_{2}F_{1}\left(Q_{3}\right)={}_{2}F_{1}\left(1,k_{\text{a}}+k_{\text{d}};k_{\text{d}}+1;\frac{J_{1}-1}{J_{1}-J_{0}}\right),
\]
\[
_{2}F_{1}\left(Q_{4}\right)={}_{2}F_{1}\left(1,k_{\text{a}}+k_{\text{d}};k_{\text{d}}+1;\frac{J_{1}}{J_{1}-J_{0}}\right).
\]

The probability $P_{A}$ can also be interpreted as the Promoter activity
(\% maximal response), defined as the fraction of cells in a cell
population with $z>Z_{th}$ \cite{karmakar2004graded}.

\subsection{Parameter Estimation}

\label{subsec:paraest_PA}

\begin{figure}[H]
\begin{centering}
\subfloat[]{\begin{centering}
\centering{}\includegraphics[width=6cm,height=4cm]{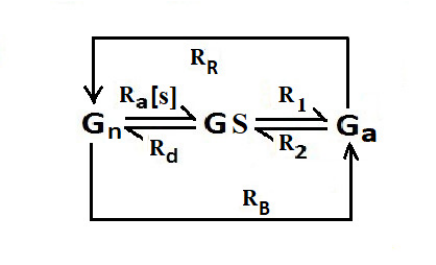}
\par\end{centering}
}\subfloat[]{\begin{centering}
\centering{}\includegraphics[width=4cm,totalheight=4cm,height=3cm]{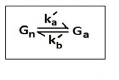}
\par\end{centering}
}
\par\end{centering}
\centering{}\subfloat[]{\begin{centering}
\centering{}\qquad{}\includegraphics[width=8cm,totalheight=4cm,height=6cm]{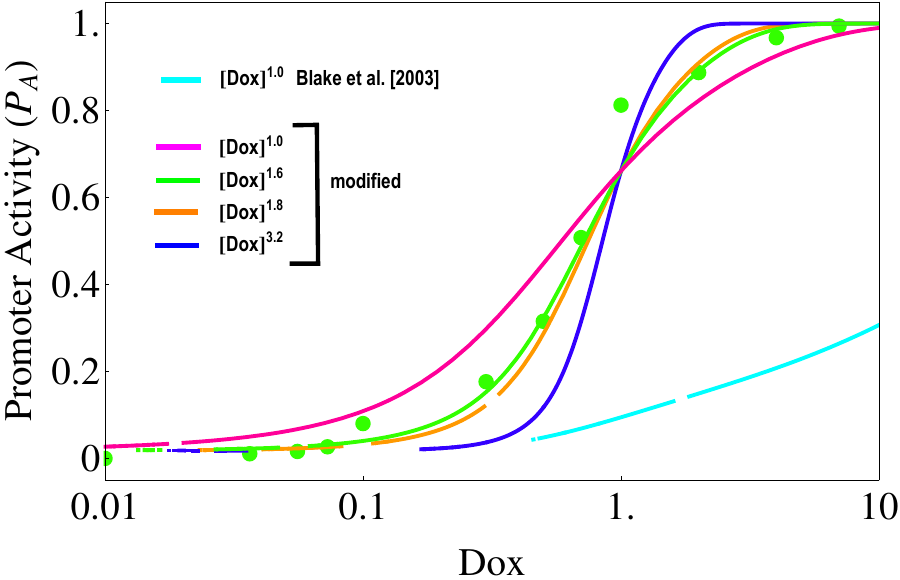}
\par\end{centering}
}\caption{{\small (a) Intermediate state assumption (b) equivalent transition
(c) parameter prediction: curves for Dox power = 1.0 (cyan, Blake
et al. form \cite{blake2003noise}), 1.0 (magenta, modified form),
1.6 (green, modified form), 1.8 (orange, modified form) and 3.2 (blue,
modified form).}}
\label{fig:Intermediate-state-assumption} 
\end{figure}

In this section we aim to find the best-fitted parameters for the
reaction scheme \ref{fig:2-Two-state-activator}a. In order to do
so, we put aside the concept of the Hill function and try to form
the GAL-dependent parameters used by Blake et al. \cite{blake2003noise}.
First of all, we theoretically determine the form of the parameters
that are responsible for stochasticity in that circuit. In this context,
we assume there exists an intermediate state called `gene-dox conjugate
state' ($GS$) between normal state $G_{n}$ and active state $G_{a}$
(Fig.~\ref{fig:Intermediate-state-assumption}a). The transition
rate constant from $G_{n}$ to $GS$ carries doxycycline molecule
{[}$S],$ that binds to the gene. Then, with the help of some enzyme
(stimuli) or inducer, the gene gets activated ($G_{a})$ via rate
constant $R_{1}.$ The reaction rates $R_{d}$ and $R_{2}$ are the
reverse transitions of $R_{a}$ and $R_{1}$, respectively. There
exists a direct forward transition ($R_{B}$) from $G_{n}$ to $G_{a}$
and a direct reverse transition ($R_{R})$ from $G_{a}$ to $G_{n}$.

The kinetic equations for the reaction scheme in Fig.~\ref{fig:Intermediate-state-assumption}a
can be written as (the terms within the parentheses denote molecular
concentrations):

\begin{subequations}
\begin{equation}
\frac{d[G_{a}]}{dt}=R_{1}[GS]+R_{B}[G_{n}]-\left(R_{2}+R_{R}\right)[G_{a}],
\end{equation}

\begin{equation}
\frac{d[GS]}{dt}=R_{a}[S^{m}][G_{n}]-R_{d}[GS]+R_{2}[G_{a}]-R_{1}[GS],
\end{equation}

\begin{equation}
\frac{d[G_{n}]}{dt}=R_{d}[GS]+R_{R}[G_{a}]-\left(R_{a}[S^{m}]+R_{B}\right)[G_{n}],
\end{equation}
\end{subequations}

\begin{equation}
[G_{n}]+[GS]+[G_{a}]=1.\label{eq:inter state-4}
\end{equation}

Applying the steady state conditions, $\frac{d[G_{a}]}{dt}=0,\frac{d[GS]}{dt}=0,\frac{d[G_{n}]}{dt}=0$,
and solving we get,

\begin{equation}
[G_{a}]=\frac{k_{a}^{'}}{k_{a}^{'}+k_{d}^{'}},\label{eq:inter state solution}
\end{equation}
where, 
\begin{equation}
k_{a}^{'}=R_{1}[R_{a}[S^{m}]+\frac{R_{B}}{R_{1}}R_{d}+R_{B}],\label{eq:equivalent k1}
\end{equation}

\begin{equation}
k_{b}^{'}=R_{2}^{c}[R_{a}[S^{m}]+R_{d}+R_{c}],\label{eq:equivalent k2}
\end{equation}
with $R_{2}^{c}=(R_{2}+R_{R})$ and $R_{c}=\frac{R_{1}R_{R}+R_{2}R_{B}}{R_{2}^{c}}$.\\
 At this stage of analysis, we further assume that $R_{R}$ completely
unwinds the dox from the gene and hence is a function of $[S]$ such
as $R_{R}=\frac{\alpha}{[S^{m}]}$, $\alpha$ being a proportionality
constant.

When the intermediate state $GS$ is absent, $G_{a}$ takes the same
form as in Eq.~(\ref{eq:inter state solution}). Additionally, from
the Eq.~(\ref{eq:equivalent k1}) and (\ref{eq:equivalent k2}) we
may obtain the exact form of the GAL (dox) dependent rate constants
$k_{a}$ and $k_{d}$. In \cite{blake2003noise}, Blake et al. used
the GAL-dependent parameters as $k_{a}=0.2\times GAL+0.02$ and $k_{d}=0.1\times GAL+0.01+0.07/GAL+0.007/GAL$.
We have tried these values and found a curve (cyan in Fig.~\ref{fig:Intermediate-state-assumption}c)
nowhere near the experimental data, even though it follows the similar
sigmoid nature at much higher GAL concentrations. Therefore we have
modified the parameter values a bit and put a power $(m\geq1)$ on
dox (or GAL) to make them more general. In order to explain the result
found in the experiment performed by Rossi et al. \cite{rossi2000transcriptional},
we keep the power of dox as 1.6 for the activator-only model and 1.8
for the repressor-only model. Then, we show that the power goes near
to 3.2 when both activator and repressor mutually acts competitively
in the circuit. By choosing different numerical values, we have been
able to fit the experimental data with theoretical curves quite nicely
(Fig.~\ref{fig:2-Two-state-activator}b). We have checked by minimizing
the relative error and mean squared error to support the robustness
of our parameter estimations. The value of parameters for the activator-only
model are: $k_{a}=1.2\times S^{1.6}+0.2$; $k_{d}=0.01\times S^{1.6}+0.001+0.0279/S^{1.6}$;
$J_{m}\in[5-250]$; $J_{0}=0.0001$; $k_{m}=1$, and $Z_{th}=0.987.$

\section{Two-state model where repressor acts only}

\label{sec:twostatrep}

\begin{figure}[H]
\begin{centering}
\subfloat[]{\begin{centering}
\centering{}\includegraphics[width=6cm,totalheight=4cm,height=4cm]{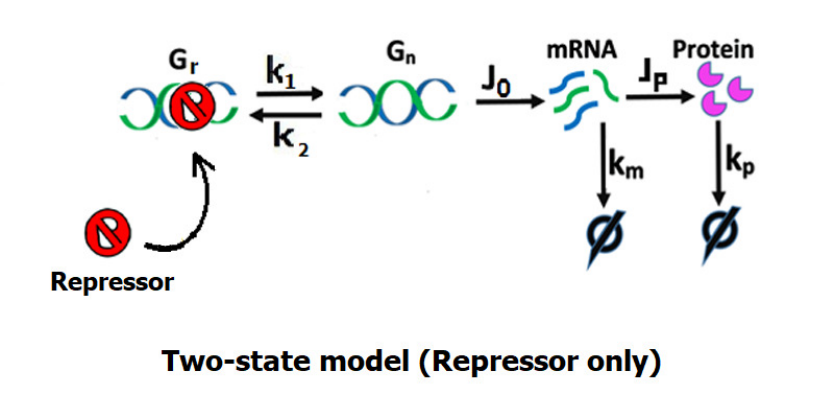}
\par\end{centering}
}\subfloat[]{\begin{centering}
\centering{}\includegraphics[width=6cm,totalheight=4cm,height=4cm]{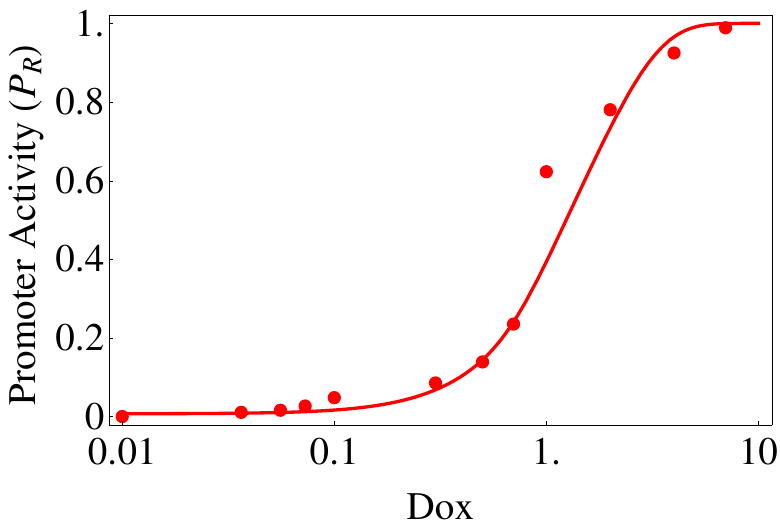}
\par\end{centering}
}
\par\end{centering}
\caption{{\small (a) Two-state model where repressor acts solely and inhibits
the transcription (b) Dose-response curve (solid curve) fitted with
experimental data (solid circles) when only repressor is present (activator
absent) in the promoter of gene.}}
\label{fig:Two-state repressor-model} 
\end{figure}

We now focus our attention to a two-state repressor only model (Fig.~\ref{fig:Two-state repressor-model}a)
where mRNA is produced spontaneously from normal state of gene $G_{n}$
with a very low leaky basal rate $J_{0}$ but no mRNA is being produced
when the gene is repressed by some TFs, then the gene is assumed to
be OFF $(G_{r}).$ Whenever a repressor molecule gets attached to
the promoter it effectively inhibits the transcription. We may calculate
the promoter activity for that model as 
\begin{eqnarray}
P(z>Z_{th}) & = & 1-\frac{\intop_{0}^{Z_{th}}P(z)\ dz}{\intop_{0}^{1}P(z)\ dz},\nonumber \\
 & = & 1-\frac{Z_{\text{th}}^{\frac{k_{1}}{k_{m}}}\,_{2}F_{1}\left(\frac{k_{1}}{k_{m}},1-\frac{k_{2}}{k_{m}};\frac{k_{1}}{k_{m}}+1;\frac{k_{1}Z_{\text{th}}}{k_{1}+k_{2}}\right)}{\,_{2}F_{1}\left(\frac{k_{1}}{k_{m}},1-\frac{k_{2}}{k_{m}};\frac{k_{1}}{k_{m}}+1;\frac{k_{1}}{k_{1}+k_{2}}\right)}=P_{R}\thinspace(\mbox{say}).\thinspace
\end{eqnarray}

Interestingly, by using parameters: $k_{1}=0.02+1.8\times S^{1.8}$;\,$k_{2}=0.001+0.14\times S^{1.8}+0.097/S^{1.8}$~;
$J_{0}=0.0001;\thinspace k_{m}=1$ and $Z_{th}=0.70$, our theoretical
plot further agrees well with repression (Fig.~\ref{fig:Two-state repressor-model}b)
at a particular dox concentration (\textit{i.e.} $S^{1.8})$ found
in \cite{rossi2000transcriptional}. By applying gene-dox complex
formation, we have estimated these parameter values following the
same line of analysis as in the previous section.

\section{Competitive regulatory architecture}

\label{sec:competitive}

\begin{figure}[H]
\begin{centering}
\subfloat[]{\begin{centering}
\centering{}\includegraphics[width=8cm,totalheight=4cm,height=5cm]{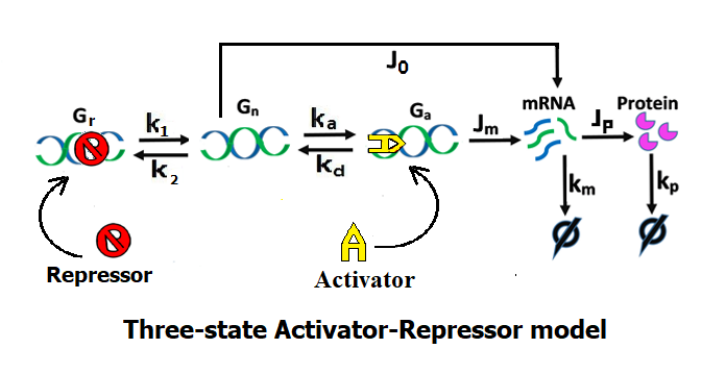}
\par\end{centering}
}
\par\end{centering}
\centering{}\subfloat[]{\begin{centering}
\centering{}\includegraphics[width=10cm,totalheight=5cm,height=6cm]{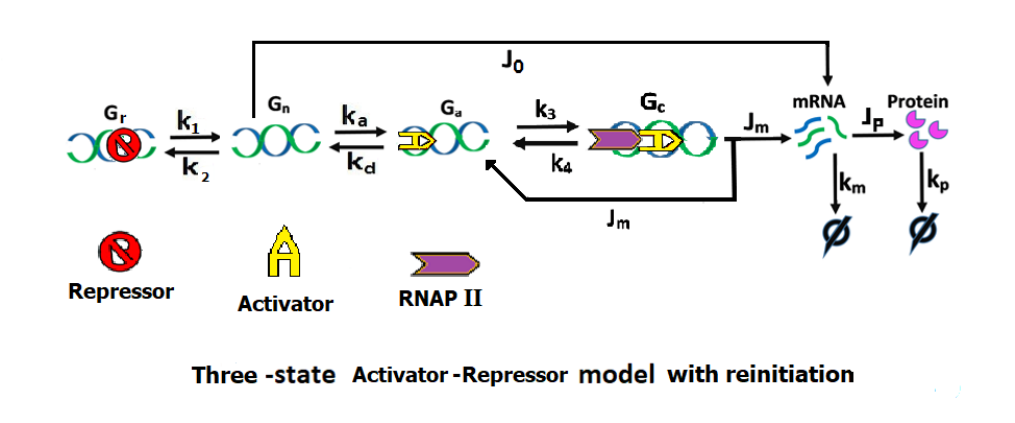}
\par\end{centering}
}\caption{{\small Competitive binding of activator and repressor model (a) without
reinitiation, (b) with reinitiation.}}
\label{fig:competitive-binding} 
\end{figure}

There are experimental studies on an eukaryotic transcriptional regulatory
network where the activators and repressors bind the same sites of
the promoter mutually exclusively \cite{rossi2000transcriptional}.
The important feature of the experimental observation is that the
activator and repressor concentrations are controlled by the single
inducer doxycycline (dox). By varying the dox, the activator and repressor
concentrations are varied simultaneously in the experiment. The authors
observed an all-or-none response when a combination of activators
and repressors act on the same promoter, whereas either alone shows
a graded response. The all-or-none and the graded response of an activator-repressor
system are explained by considering three states of a gene mainly,
inactive ($G_{r}$), normal ($G_{n}$) and active ($G_{a}$), where
the activator and repressor compete with each other to bind to the
normal state, which is an open state (Fig.~\ref{fig:competitive-binding}a)
\cite{karmakar2010conversion}. A successful binding of the activator
(repressor) turns the normal state into the active (inactive or repressed)
state. Random transitions take place due to random binding and unbinding
events of activators and repressors.

In our present analysis we consider the reinitiation of transcription
by RNAP II along with the competitive binding events of activators
and repressors. The detailed reaction scheme is shown in Fig.~\ref{fig:competitive-binding}b.

The promoter activity calculated for the activator-repressor competitive
system as 
\begin{equation}
\begin{aligned}P(z>Z_{th}) & =1-\frac{\intop_{0}^{Z_{th}}P(z)\ dz}{\intop_{0}^{1}P(z)\ dz},\\
 & =1-\frac{Z_{\text{th}}^{\frac{k_{\text{ON}}}{k_{m}}}{}_{2}F_{1}\left(1-\frac{k_{\text{OFF}}}{k_{m}},\frac{k_{\text{ON}}}{k_{m}};\frac{k_{\text{ON}}}{k_{m}}+1;\frac{k_{\text{ON}}Z_{\text{th}}}{k_{\text{OFF}}+k_{\text{ON}}}\right)}{_{2}F_{1}\left(1-\frac{k_{\text{OFF}}}{k_{m}},\frac{k_{\text{ON}}}{k_{m}};\frac{k_{ON}}{k_{m}}+1;\frac{k_{\text{ON}}}{k_{\text{OFF}}+k_{\text{ON}}}\right)},\\
 & =P_{AR}\ (say),\,\\
\\\end{aligned}
\label{eq:promact_AR}
\end{equation}
where, $_{2}F_{1}(a;b;z)$ is a Hypergeometric function.

\subsection{Parameter estimation for Activator-Repressor system}

\label{subsec:para_est_PAR}

\begin{figure}[H]
\begin{centering}
\subfloat[]{\begin{centering}
\centering{}\includegraphics[width=6cm,totalheight=5cm,height=4cm]{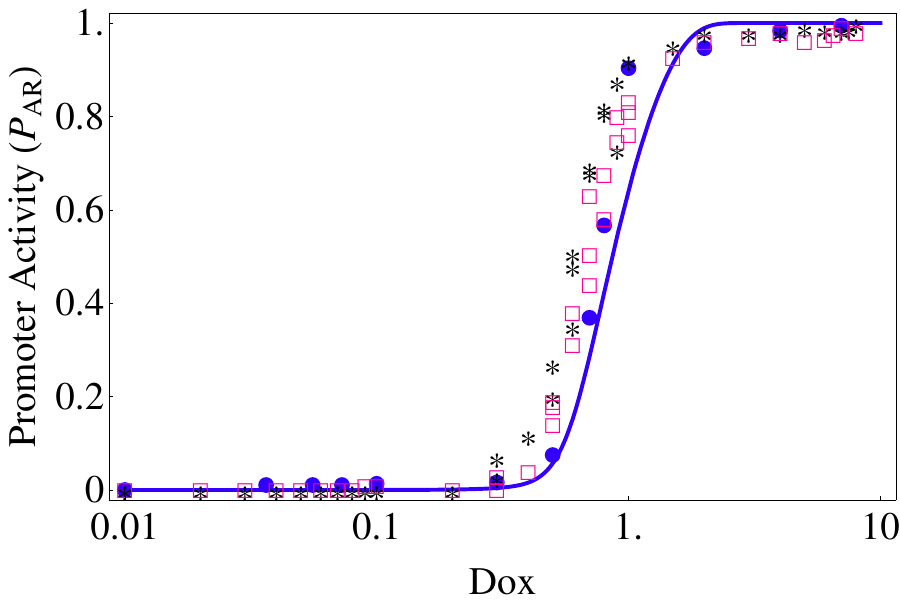}
\par\end{centering}
}\subfloat[]{\begin{centering}
\centering{}\includegraphics[width=6cm,totalheight=5cm,height=4cm]{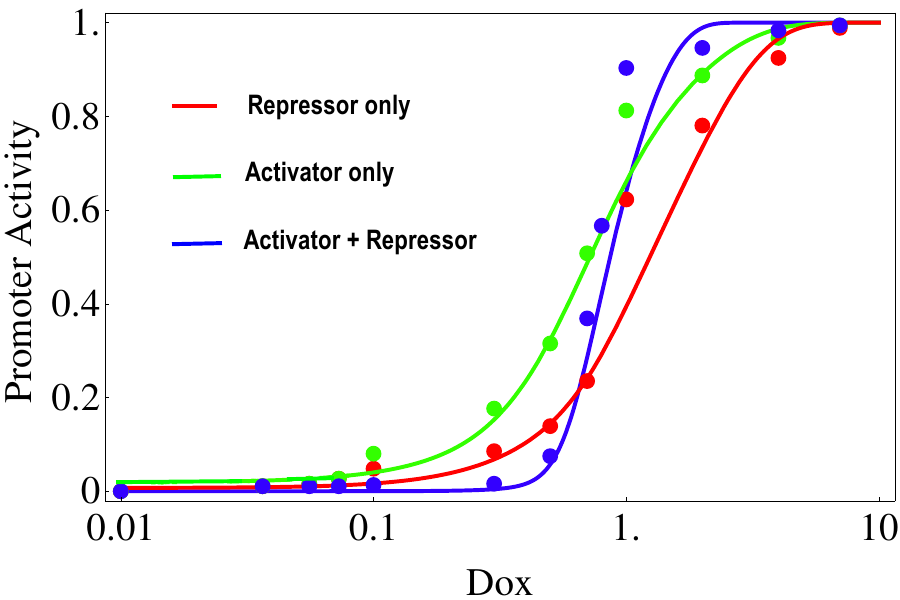}
\par\end{centering}
}
\par\end{centering}
\centering{}\subfloat[]{\begin{centering}
\centering{}\includegraphics[width=6cm,totalheight=5cm,height=4cm]{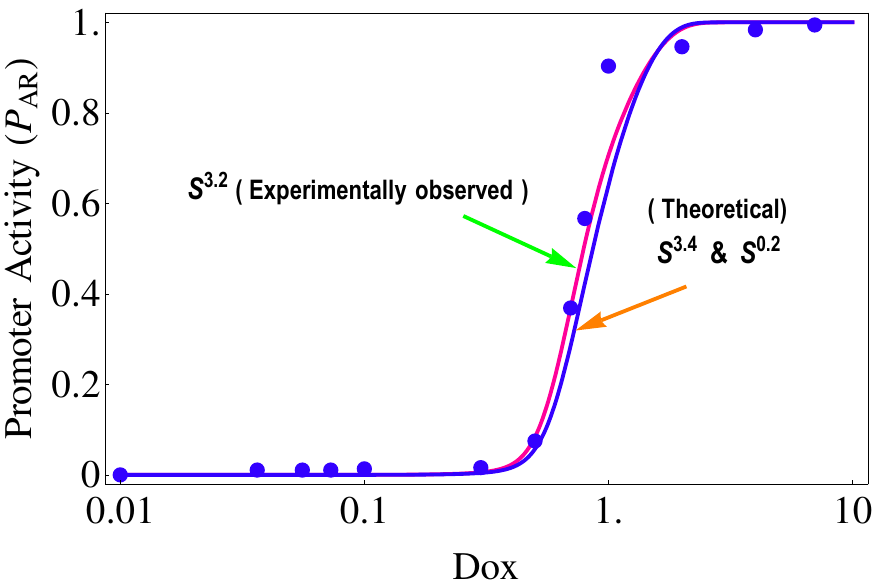}
\par\end{centering}
}\subfloat[]{\begin{centering}
\includegraphics[width=6cm,totalheight=4cm,height=4cm]{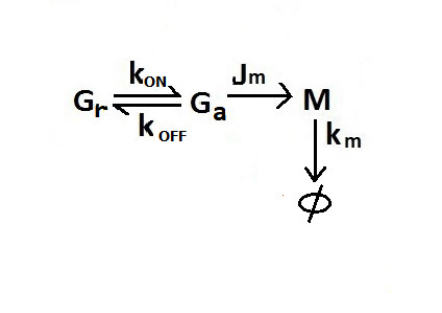} 
\par\end{centering}
}\caption{{\small (a) Dose-response curve (solid curve) fitted with experimental
data (solid circles) when both activator and repressor compete to
sit on the promoter of the gene. Pink squares (Black stars) are obtained
by simulation based on the Gillespie algorithm \cite{gillespie1977exact}
corresponding to the reaction scheme \ref{fig:competitive-binding}a
(\ref{fig:competitive-binding}b) using $k_{3}=350.0$ and $k_{4}=70.0$
(b) Dose-response curve (solid curves: green- activator-only, red-
repressor-only and blue- activator-repressor in competition) fitted
with experimental data (solid circles:~green- activator-only, red-
repressor-only and blue- activator-repressor in competition). (c)
Dose-response curve for activator-repressor showing a close agreement
of theoretically obtained curve with dox power combination of 3.4
\& 0.2 (blue curve) with experimentally found value of 3.2 (pink curve).
(d) Two-state equivalent simplified circuit of the three-state activator-repressor
competitive model.}}
\label{fig:AR Curve-fitting_all} 
\end{figure}

In this section, we attempt to find the possible set of parameters
that fits the experimental data of Ref.~\cite{rossi2000transcriptional}.
They have mentioned in their paper that the dose-response curve obtained
from their experiment might follow the Hill function with different
Hill coefficients. For different Hill coefficients the model behaves
as repressor-only model or activator-only model and a competitive
model where both activator and repressor are in competition for binding
to the same region of the promoter of gene. In our investigation,
we find that the parameter set is much similar to those mentioned
by \cite{blake2003noise} rather than the Hill function. Although
in \cite{blake2003noise} the authors used GAL as activator with power
$m=1.0$ but here we keep the powers of dox (or GAL) as 1.8 for repressor
and 1.6 for activator-only following the Hill coefficient values found
in \cite{rossi2000transcriptional} (Fig.~\ref{fig:AR Curve-fitting_all}a
and Fig.~\ref{fig:AR Curve-fitting_all}b). In order to explore the
competitive model we consider a simplified equivalent two-state model
(fig. \ref{fig:AR Curve-fitting_all}d) of reaction scheme \ref{fig:competitive-binding}a.
Here, we neglect $J_{0}$ in comparison with $J_{m}\ or\ J_{1}$ (as
$J_{0}<<J_{1})$.

This assumption does not make any qualitative change in PDF which
we have checked explicitly \cite{karmakar2010conversion}. Moreover,
it also holds true in case of mean mRNA and noise strength. It has
been known for quite sometime that, the deterministical determination
of probability for a three state activator-repressor competitive binding
model is very difficult when basal rate $J_{0}$ is involved \cite{karmakar2010conversion}.
The deterministic reaction rate equations for reaction scheme shown
in fig.\ref{fig:competitive-binding}a are given by

\begin{subequations}
\begin{equation}
\frac{d[G_{r}]}{dt}=k_{2}[G_{n}]-k_{1}[G_{r}],\label{eq:3-state-1}
\end{equation}
\begin{equation}
\frac{d[G_{n}]}{dt}=k_{d}[G_{a}]-k_{a}[G_{n}],\label{eq:3-state-2}
\end{equation}
\begin{equation}
\frac{d[G_{a}]}{dt}=k_{a}[G_{n}]-k_{d}[G_{a}],\label{eq:3-state-3}
\end{equation}
\begin{equation}
\frac{d[M]}{dt}=J_{1}[G_{a}]+J_{0}[G_{n}]-k_{m}[M],\label{eq:3-state-4}
\end{equation}
\end{subequations}

\begin{equation}
[G_{n}]+[G_{r}]+[G_{a}]=1.\label{eq:3-state-5}
\end{equation}
In steady state, $\frac{d[M]}{dt}=0$ and we have 
\begin{equation}
[M]=\frac{J_{1}[G_{a}]+J_{0}[G_{n}]}{k_{m}}\simeq\frac{J_{1}}{k_{m}}[G_{a}],\qquad(J_{0}<<J_{1}).\label{eq:3-state meanM}
\end{equation}

Applying the steady state condition, $\frac{d[G_{a}]}{dt}=0,\frac{d[G_{r}]}{dt}=0,\frac{d[G_{n}]}{dt}=0$
and solving, we get 
\begin{equation}
[G_{a}]=\frac{k_{1}k_{a}}{k_{1}k_{a}+k_{d}(k_{1}+k_{2})}.\label{eq:3- state-sol-1}
\end{equation}
Therefore, Eq.\,\ref{eq:3-state meanM} gives 
\begin{equation}
[M]=\frac{J_{1}}{k_{m}}\frac{k_{1}k_{a}}{k_{1}k_{a}+k_{d}(k_{1}+k_{2})}.\label{eq:3-state-sol-2}
\end{equation}

Now, the reaction rate equations for the simplified equivalent model
(see reaction scheme in Fig.~\ref{fig:AR Curve-fitting_all}d) are

\begin{subequations}
\begin{equation}
\frac{d[G_{r}]}{dt}=k_{OFF}[G_{a}]-k_{ON}[G_{r}],\label{eq:equiv 2-state}
\end{equation}
\begin{equation}
\frac{d[M]}{dt}=J_{m}[G_{a}]-k_{m}[M].\label{eq:equiv 2-state-2}
\end{equation}
\begin{equation}
\begin{aligned}[][G_{r}]+[G_{a}] & =1.\end{aligned}
\label{eq:equiv 2-state-3}
\end{equation}
\end{subequations}
 The steady state condition gives 
\begin{equation}
[M]=\frac{J_{m}}{k_{m}}\frac{k_{ON}}{(k_{ON}+k_{OFF})},\label{eq:equiv 2-state sol}
\end{equation}
we now write $J_{m}=J_{1}$ and compare Eq.~(\ref{eq:3-state-sol-2})
with Eq.~(\ref{eq:equiv 2-state sol}) we get 
\begin{eqnarray}
k_{ON} & = & k_{1}k_{a},\label{eq: k-on}\\
k_{OFF} & = & k_{d}(k_{1}+k_{2}).\label{eq:k-off}
\end{eqnarray}
Notice that, we get a form of the parameters that have a dominant
role over the others for the system. We see that algebric composition
of the component-parameters like $k_{1},k_{a},k_{d},k_{2}$, etc.
to form $k_{ON}$ and $k_{OFF}$ actually leads to an addition of
powers of dox molecule. Theoretically, we obtain the power of dox
(S) for the activator-repressor competitive system as 3.4 which has
a close agreement with the experimentally observed value 3.2 (Fig.~\ref{fig:AR Curve-fitting_all}c).
Here, by taking the dox-dependent parts of the component-parameters
and using Eq.~(\ref{eq: k-on}) and Eq.~(\ref{eq:k-off}) we get,
$k_{ON}=1.08\times S^{3.4}$ and $k_{OFF}=0.0032\times S^{3.4}+0.0027/S^{3.4}+0.0089\times S^{0.2}+0.00097/S^{0.2}$.
On top of that, while working with the parameter estimation we notice
that besides the power of dox, there are other parameters like $Z_{th}$
that affect the slope of the curves. In our analysis we can put $Z_{th}$
at any value between 70\% to 99\% to find a best fit curve. We also
perform a simulation (shown in Fig.\medspace{}\ref{fig:AR Curve-fitting_all}a),
based on the Gillespie algorithm \cite{gillespie1977exact} corresponds
to the reaction scheme of Fig.\medspace{}\ref{fig:competitive-binding}a
(Pink squares) and Fig.\medspace{}\ref{fig:competitive-binding}b
(black stars) respectively, which gives a good agreement to the experimental
and as well as the theoretical results. Moreover, to show the robustness
of our parameter estimation we examine the sensitivity against the
parameters and minimize the standard errors for each data points by
using chi-square fitting method (see Appendix-B).

\subsection{Stochastic Analysis}

\label{subsec:stochastic} Let, there are $l$ copy numbers of a particular
gene exist in the cell. We consider the reaction scheme in Fig.~\ref{fig:competitive-binding}b
and let $P(n_{1},n_{2},n_{3},n_{4},n_{5},t)$ be the probability that
at time t, there are $n_{4}$ number of mRNAs and $n_{5}$ number
of proteins molecules with $n_{3}$ number of genes in the initiation
complex ($G_{c}$), $n_{2}$ number of genes in the active state ($G_{a}$),
and $n_{1}$ number of genes in the normal state ($G_{n}$). The number
of gene in the inactive states ($G_{r}$) are thus ($l-n_{1}-n_{2}-n_{3}$).
The time evolution of the probability (assuming $J_{0}=0)$ is given
by 
\begin{equation}
\begin{split}\frac{\partial P(n_{i},t)}{\partial t} & =k_{1}[\{l-(n_{1}-1+n_{2}+n_{3})\}P(n_{1}-1,n_{2},n_{3},n_{4},n_{5},t)\\
 & -\{l-(n_{1}+n_{2}+n_{3})\}P(n_{i},t)]\\
 & +k_{2}[(n_{1}+1)P(n_{1}+1,n_{2},n_{3},n_{4},n_{5},t)-n_{1}P(n_{i},t)]\\
 & +k_{a}[(n_{1}+1)P(n_{1}+1,n_{2}-1,n_{3},n_{4},n_{5},t)-n_{1}P(n_{i},t)]\\
 & +k_{d}[(n_{2}+1)P(n_{1}-1,n_{2}+1,n_{3},n_{4},n_{5},t)-n_{2}P(n_{i},t)]\\
 & +k_{3}[(n_{2}+1)P(n_{1},n_{2}+1,n_{3}-1,n_{4},n_{5},t)-n_{2}P(n_{i},t)]\\
 & +k_{4}[(n_{3}+1)P(n_{1},n_{2}-1,n_{3}+1,n_{4},n_{5},t)-n_{3}P(n_{i},t)]\\
 & +J_{m}[(n_{3}+1)P(n_{1},n_{2}-1,n_{3}+1,n_{4}-1,n_{5},t)-n_{3}P(n_{i},t)]\\
 & +k_{m}[(n_{4}+1)P(n_{1},n_{2},n_{3}n_{4}+1,n_{5},t)-n_{4}P(n_{i},t)]\\
 & +J_{p}[n_{4}P(n_{1},n_{2},n_{3},n_{4},n_{5}-1,t)-n_{4}P(n_{i},t)]\\
 & +k_{p}[(n_{5}+1)P(n_{1},n_{2},n_{3},n_{4},n_{5}+1,t)-n_{5}P(n_{i},t)],
\end{split}
\label{eq:Competitive stoch Prob}
\end{equation}
where, $i=1,2,\dots,5$.

~We can derive the mean, variance and the Fano factor of mRNAs and
proteins from the moments of Eq. (\ref{eq:Competitive stoch Prob})
with the help of a generating function (Appendix-A). The mean mRNA
and protein are given by \\
 
\begin{equation}
m^{CWR}=\frac{J_{m}k_{1}k_{a}k_{3}}{l_{3}k_{m}};\quad\quad p^{CWR}=\frac{m^{CWR}\,J_{p}}{k_{p}}.\label{eq:CWR meanM and meanP}
\end{equation}
The expression for the Fano factors at mRNA and protein levels are
given by 
\begin{equation}
FF_{m}^{CWR}=1-m^{CWR}-\frac{J_{m}k_{3}(h_{2}k_{m}+k_{1}k_{a})}{(k_{3}(h_{2}(J_{m}+k_{4})-k_{1}k_{a})+h_{8}(k_{a}(k_{d}-k_{1})-h_{2}h_{6}))k_{m}},\label{eq:CWR FFm}
\end{equation}
\begin{equation}
FF_{p}^{CWR}=1-p^{CWR}+\frac{J_{p}J_{m}k_{3}(k_{m}k_{p}-k_{1}k_{a})}{l_{10}}+\frac{J_{p}J_{m}k_{3}(h_{2}k_{m}+k_{1}k_{a})}{l_{12}}+\frac{J_{p}}{k_{m}+k_{p}},\label{eq:CWR FFp}
\end{equation}
where 
\[
\begin{aligned}l_{12} & =l_{1}l_{10}(k_{1}k_{a}(k_{3}+l_{8})-k_{3}l_{2}J_{m}-k_{3}k_{4}l_{2}-k_{a}k_{d}l_{8}+l_{6}l_{8}l_{2});\\
l_{11} & =l_{10}+l_{1}(k_{3}J_{m}-k_{1}k_{a}+k_{a}k_{d}+k_{3}k_{4}-l_{4}l_{5}-l_{4}l_{8}-l_{6}l_{8};\\
l_{10} & =(k_{3}l_{9}+l_{7}(k_{a}k_{d}-k_{a}k_{1}-l_{4}l_{5}))k_{p}(k_{m}+k_{p});\\
l_{9} & =(l_{4}(J_{m}+k_{4})-k_{1}k_{a});\quad l_{8}=J_{m}+k_{4}+k_{m};\quad l_{7}=J_{m}+k_{4}+k_{p};\\
l_{6} & =k_{d}+k_{3}+k_{m};\quad l_{5}=k_{d}+k_{3}+k_{p};\quad l_{4}=k_{1}+k_{2}+k_{a}+k_{p};\\
l_{3} & =k_{1}k_{a}k_{3}+(k_{1}k_{d}+k_{2}k_{d}+k_{1}k_{a})(J_{m}+k_{4});\\
l_{2} & =k_{1}+k_{2}+k_{a}+k_{m};\quad l_{1}=(k_{m}+k_{p})k_{m}k_{p}.
\end{aligned}
\]

For the without-reinitiation scheme (Fig.~\ref{fig:competitive-binding}a),
the required expression for the mean level of mRNA ($m^{CWTR}$) and
protein ($p^{CWTR}$) are given by 
\begin{equation}
m^{CWTR}=\frac{J_{m}k_{1}k_{a}}{r_{1}k_{m}};\quad\quad p^{CWTR}=\frac{m^{CWTR}\,J_{p}}{k_{p}}.\label{eq:CWTR meanM_meanP}
\end{equation}
The corresponding Fano factors at mRNA ($FF_{m}^{CWTR}$) and protein
($FF_{p}^{CWTR}$) level are given by 
\begin{equation}
FF_{m}^{CWTR}=1-m^{CWTR}+\frac{J_{m}(r_{5}k_{m}+k_{1}k_{a})}{r_{2}k_{m}},\label{eq:CWTR FFm}
\end{equation}
\begin{equation}
FF_{p}^{CWTR}=1-p^{CWTR}+X\label{eq:CWTR FFp}
\end{equation}
where, $X=\frac{J_{p}}{k_{m}+k_{p}}+\frac{J_{p}J_{m}k_{1}k_{a}}{r_{9}k_{p}(k_{m}+k_{p})}+\frac{J_{p}J_{m}(r_{6}k_{m}+r_{9})}{r_{9}(k_{m}+k_{d})(k_{m}+k_{p})}+\frac{J_{p}J_{m}k_{a}k_{d}(k_{m}+k_{1})r_{10}}{r_{4}(k_{m}+k_{d})(k_{m}+k_{p})k_{m}};$
\[
r_{10}=r_{1}+r_{3}k_{m}+r_{7}k_{p};\quad r_{9}=r_{6}(k_{p}+k_{d})+k_{a}(k_{1}-k_{d});\quad r_{8}=r_{6}+k_{d};
\]
\[
r_{7}=r_{8}+k_{m};\quad r_{7}=r_{8}+k_{m};\quad r_{6}=k_{1}+k_{2}+k_{a}+k_{p};\quad r_{5}=k_{1}+k_{2}+k_{a}+k_{m};
\]
\[
r_{4}=r_{2}(r_{8}k_{p}+r_{1});\quad r_{3}=r_{5}+k_{d};\quad r_{2}=r_{1}+r_{3}k_{m};\quad r_{1}=k_{1}k_{a}+k_{1}k_{d}+k_{2}k_{d}.
\]
Here, the superscript `$C$' stands to indicate the competitive binding,
`$WTR$' represents without-reinitiation and `$WR$' implies with-reinitiation
model.

The experimental values of the rate constants $k_{i}$ ($i=1,2,\dots,6$)
are not yet available for the competitive activator-repressor model
(Fig.~\ref{fig:competitive-binding}a). In the previous sections,
we tried to find a probable set of parameters that fits the experimental
data. In this way, the activator and repressor concentrations are
controlled by the single inducer doxycycline (dox, denoted by `$S$').
We notice that the theoretically obtained parameter set is quite similar
to the parameters used in \cite{blake2003noise} in a synthetic GAL1{*}
promoter in \textit{Saccharomyces cerevisiae} (commonly known as yeast),
except they have used GAL as activator and aTc-bounded tetR as repressor.
We check the behavior of the competitive system under the parameter
set used in \cite{blake2003noise} and compare the results with the
non-competitive one \cite{das2022stochastic}. The reaction rate constants
chosen from \cite{blake2003noise} for the competitive system are:
$k_{a}=0.02+0.2\times GAL$, $k_{d}=0.01+0.1\times GAL+0.077/GAL$,\,$k_{3}=50,k_{4}=10,k_{1}=10,\,k_{2}=200\times(tetR)^{2}/[1+(C_{i}\times aTc)^{4}]^{2}$,
$J_{m}=1$, $k_{m}=1$, $J_{p}=5$, $k_{p}=0.0125$, $tetR=100$,
$C_{i}=0.1$.

We now study the behavior of mean protein levels and noise strength
with respect to the inducer (GAL and aTc) and transcriptional efficiency
(refer to glossary), respectively. We note that, the mean protein
level varies with GAL and aTc in a similar fashion in the presence
and/or absence of reinitiation (Fig.~\ref{fig:competitive-binding}a)
as in the non-competitive regulatory architecture \cite{das2022stochastic}.
The variation of the noise strength with transcriptional efficiency
for fixed aTc (500 ng/ml) is again similar to the non-competitive
regulatory architecture \cite{das2022stochastic}. Interestingly,
the variation of noise strength with transcriptional efficiency for
fixed GAL (2\%) is different from that obtained from the non-competitive
regulatory architecture \cite{das2022stochastic}. We also notice
that the noise strength is maximum at the very beginning of the transcriptional
efficiency rather than the intermediate level, as shown in Figs.~2c
and 2e of \cite{das2022stochastic}.~

\begin{figure}[H]
\centering{}\subfloat[]{\begin{centering}
\centering{}\includegraphics[width=6cm,height=4cm]{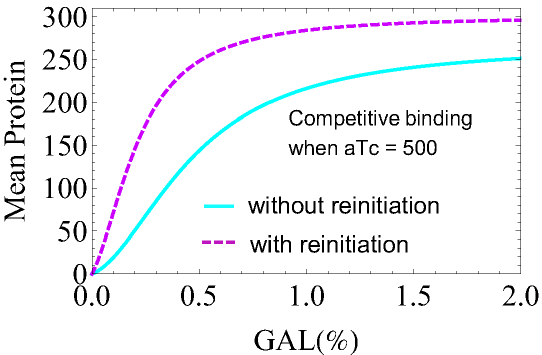}
\par\end{centering}
}\subfloat[]{\begin{centering}
\centering{}\includegraphics[width=6cm,height=4cm]{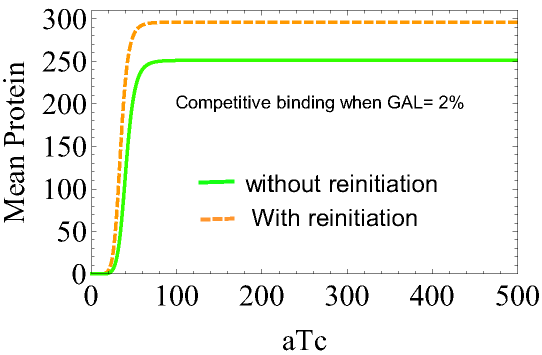}
\par\end{centering}
}\caption{{\small Variation of mean protein: (a) with GAL when aTc is fixed (b)
with aTc when GAL is fixed. Solid (dashed) lines are drawn from analytical
calculation corresponding to the reaction scheme \ref{fig:competitive-binding}a
and \ref{fig:competitive-binding}b respectively. All rate constants
are chosen from \cite{blake2003noise}.}}
\label{fig:7, MeanP vs GAL and aTc} 
\end{figure}

\begin{figure}[H]
\begin{centering}
\subfloat[]{\begin{centering}
\centering{}\includegraphics[width=6cm,height=4cm]{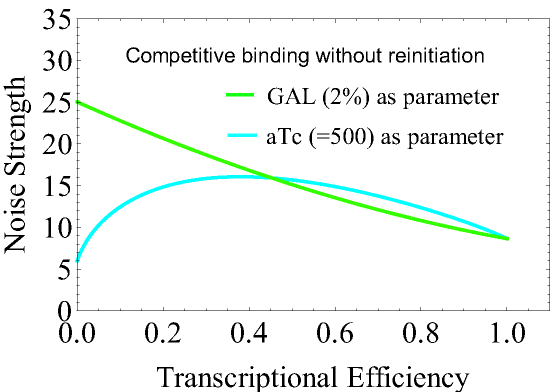}
\par\end{centering}
}\subfloat[]{\begin{centering}
\centering{}\includegraphics[width=6cm,height=4cm]{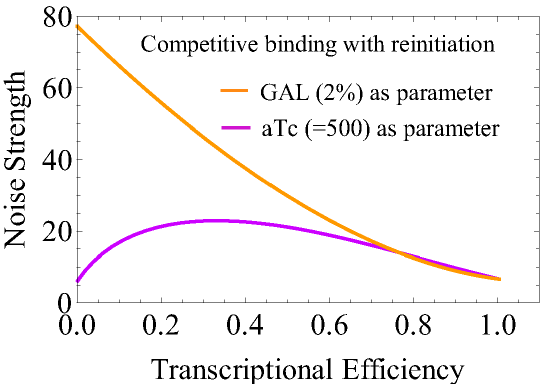}
\par\end{centering}
}\caption{{\small Variation of the noise strength with transcriptional efficiency:
(a) green (cyan) line is drawn analytically with 2\% GAL concentration
(with aTc = 500 ng/ml) from the reaction scheme \ref{fig:competitive-binding}a
(b) orange (violet) solid line is drawn analytically with 2\% GAL
concentration (with aTc = 500 ng/ml) from the reaction scheme \ref{fig:competitive-binding}b,
following the same parameter values as used by Blake et al. in \cite{blake2003noise}.}}
\par\end{centering}
\label{fig:8 noise strength vs TE} 
\end{figure}

\section{Comparison between competitive and non-competitive architecture}

\label{sec:comparison}

\begin{figure}[H]
\begin{centering}
\subfloat[]{\begin{centering}
\centering{}\includegraphics[width=9cm,totalheight=5cm,height=5cm]{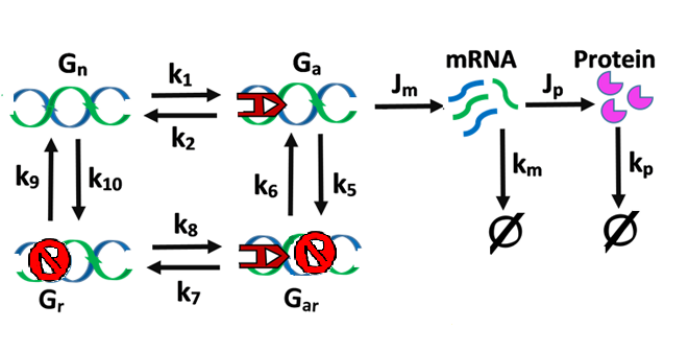}
\par\end{centering}
}
\par\end{centering}
\begin{centering}
\subfloat[]{\begin{centering}
\centering{}\includegraphics[width=9cm,totalheight=5cm,height=5cm]{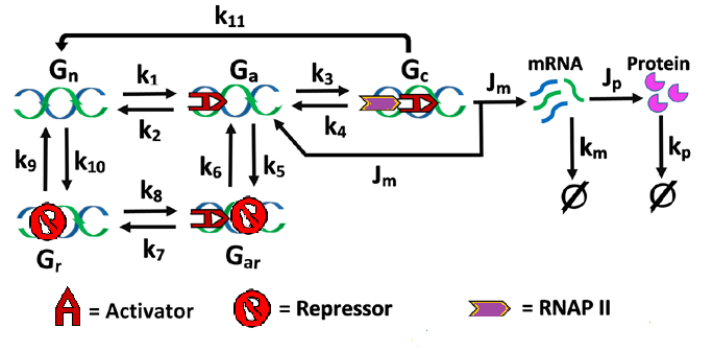}
\par\end{centering}
}
\par\end{centering}
\caption{{\small Non-competitive binding of activator and repressor (a) without
reinitiation (b) with reinitiation and with a reverse transition via
$k_{11}$.}}

\label{fig:Non-competitive-binding} 
\end{figure}

In this section, we study the nature of mean products (mRNA/Protein)
and their corresponding noise strengths (the Fano factors) for both
competitive and non-competitive binding circuits. We first note that,
the relative changes of these two physical quantities are clearly
visible for both the architectures when the reinitiation is taken
under consideration (presented by the dashed lines in characteristic
curves). Although the non-competitive circuit and its behaviour were
rigorously studied in \cite{das2022stochastic}, we put the model
diagram here to understand the comparison in a much better way. In
a competitive network, the activator and repressor compete for a common
binding region of the promoter. As a result, the gene is in either
active (ON) or repressed (OFF) state, except a normal state $(G_{n})$.
Whereas, there could be four possible genetic states: normal $(G_{n})$,
active $(G_{a})$, active-repressed $(G_{ar})$, and repressed $(G_{r})$
as shown in Fig. \ref{fig:Non-competitive-binding}a.

A reverse transition is incorporated via $k_{11}$ from the initiation
complex $(G_{c},$ arises due to reinitiation) to the normal state
$(G_{n})$ as there is a possibility of simultaneous unwinding of
RNAP-II and activator molecule, that can not be neglected \cite{das2022stochastic}.
We find that this reaction path finely affects the mean expression
and noise strength of both non-competitive circuit \cite{das2022stochastic}
and the competitive one as well.

\subsection{Role of reinitiation under the action of aTc and GAL: Anomalous behavior
at lower aTc concentration}

\label{subsec:comparirole}

In this subsection, our focus is centered around some characteristic
differences, that appear due to reinitiation when GAL (activator)
and aTc (repressor) is operating either competitively or non-competitively,
under the same reaction rate constants.

\begin{figure}[H]
\begin{centering}
\subfloat[]{\centering{}\includegraphics[width=6cm,totalheight=4cm]{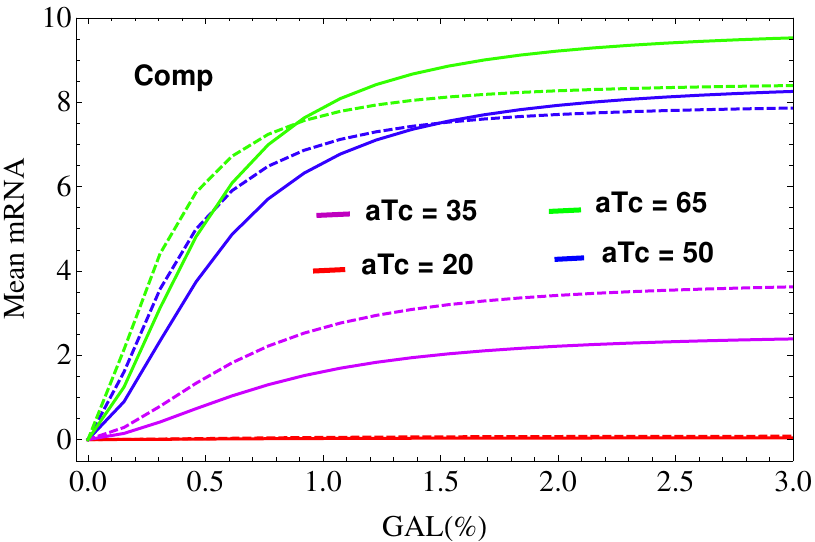}}
\subfloat[]{\centering{}\includegraphics[width=6cm,totalheight=4cm]{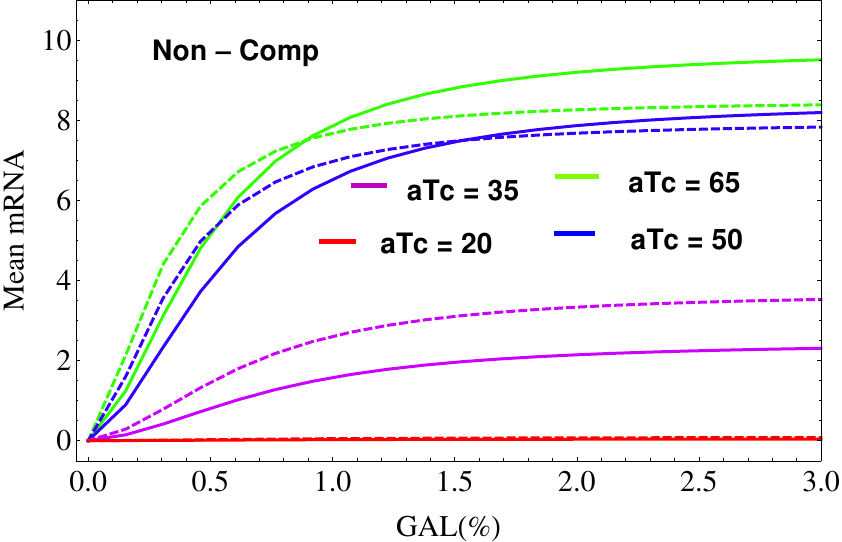}}
\par\end{centering}
\begin{centering}
\subfloat[]{\centering{}\includegraphics[width=6cm,totalheight=4cm]{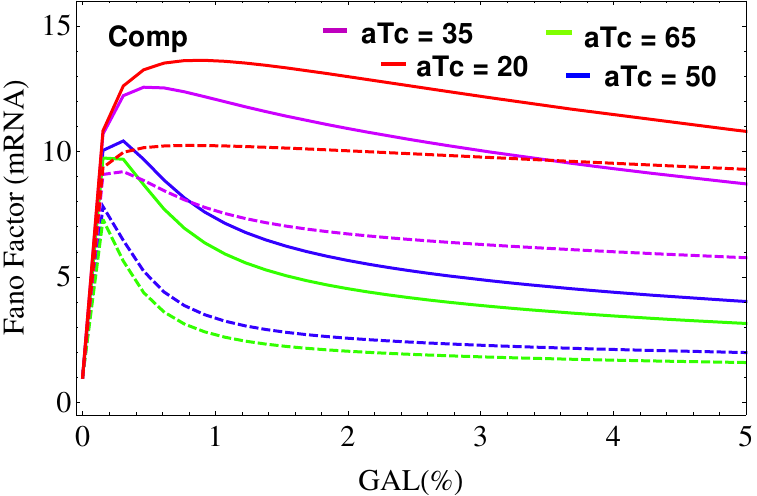}}\subfloat[]{\centering{}\includegraphics[width=6cm,totalheight=4cm]{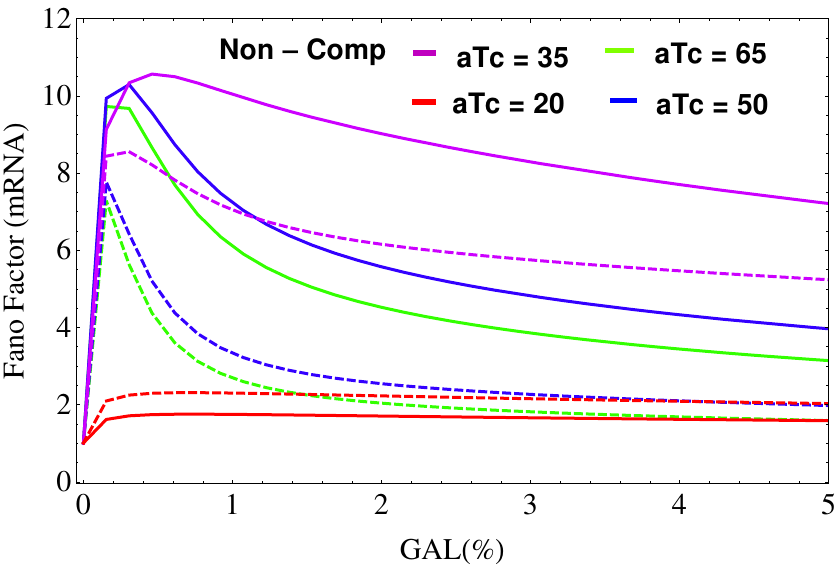}}
\par\end{centering}
\begin{centering}
\subfloat[]{\centering{}\includegraphics[width=6cm,totalheight=4cm]{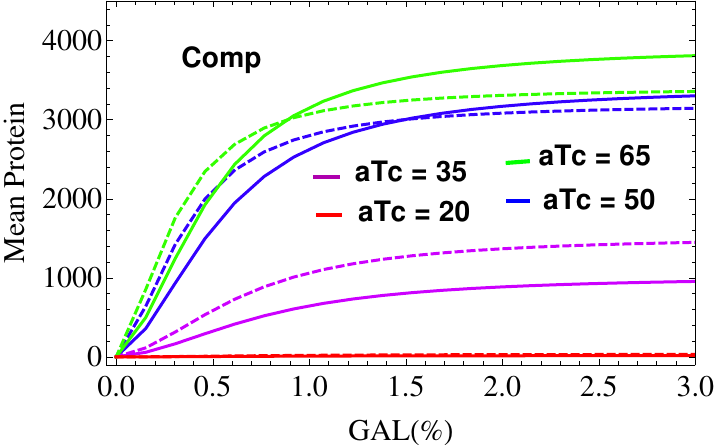}}
\subfloat[]{\centering{}\includegraphics[width=6cm,totalheight=4cm]{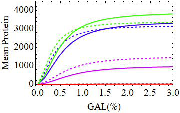}}
\par\end{centering}
\centering{}\subfloat[]{\centering{}\includegraphics[width=6cm,totalheight=4cm]{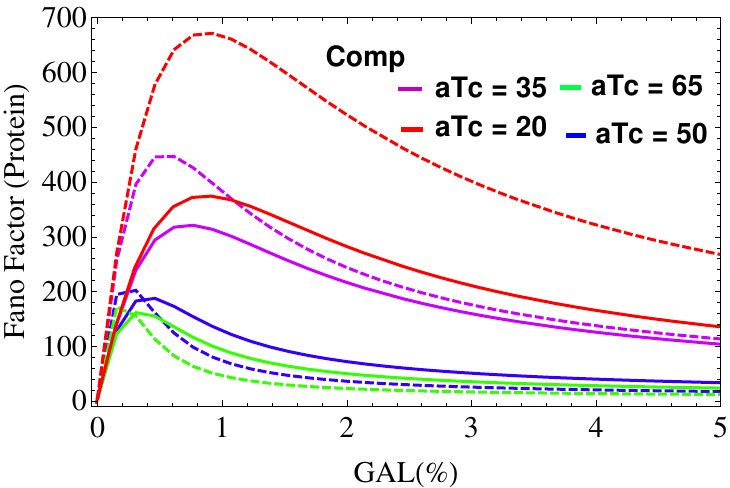}}
\subfloat[]{\centering{}\includegraphics[width=6cm,totalheight=4cm]{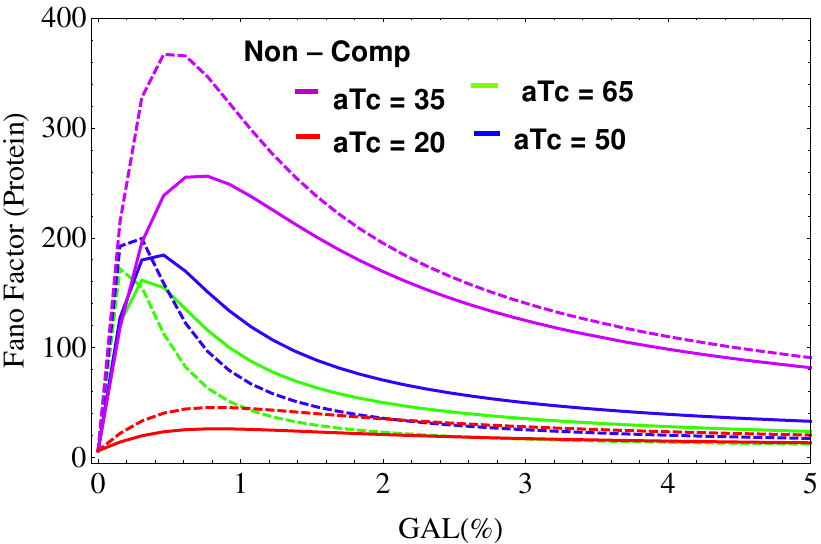}}\caption{{\small Variation of mean and Fano factor for mRNA and protein level
against GAL with different aTc. The solid (dashed) curves correspond
to the system without (with) reinitiation for reaction scheme \ref{fig:competitive-binding}a
and \ref{fig:Non-competitive-binding}a (\ref{fig:competitive-binding}b
and \ref{fig:Non-competitive-binding}b). The rate constants are chosen
from \cite{blake2003noise} with $J_{m}=15$ and $k_{11}=0$. abbreviation
'Comp (Non-comp)' implies competitive (non-competitive) model.}}\label{fig: Variation of mean and FF against GAL}
\end{figure}

In our numerical analysis, we find no observable differences in the
variation of mean mRNA as well as mean protein against GAL with different
aTc values (colour scheme: aTc = 20, red; aTc = 35, purple; atc =
50, blue; aTc = 65, green; all units are ng/ml) between both the circuits
(Figs.~\ref{fig: Variation of mean and FF against GAL}a, \ref{fig: Variation of mean and FF against GAL}b,
\ref{fig: Variation of mean and FF against GAL}e and \ref{fig: Variation of mean and FF against GAL}f).
Rather, they offer the same expression levels under identical rate
constants (chosen from \cite{blake2003noise} with $J_{m}=15$ and
$k_{11}=0$). We also notice that the noise strength (the Fano factor)
in the super-Poissonian regime ($FF_{m}>1)$ is higher in the competitive
circuit than in non-competitive one (Figs.~\ref{fig: Variation of mean and FF against GAL}c,d,g,h
and \ref{fig: Variation of mean and FF against aTc}c,d,g,h). The
findings are summarized below. 
\begin{itemize}
\item Each solid curve (presenting without-reinitiation) has a point of
intersection with the dashed curve (represents with-reinitiation)
at a particular GAL value (where, $m^{WR}=m^{WTR}$ and $p^{WR}=p^{WTR})$,
beyond which $m^{WR}\thinspace(\mbox{or,}\thinspace p^{WR})<m^{WTR}\thinspace(\mbox{or,}\thinspace p^{WTR})$.
However the curves with $aTc<45$ ng/ml $m^{WR}\thinspace(\mbox{or,}\thinspace p^{WR})>m^{WTR}\thinspace(\mbox{or,}\thinspace p^{WTR})$
always holds. 
\item The curves corresponding to aTc = 20 show some anomaly. Usually, $FF_{m}^{WTR}>FF_{m}^{WR}$
for all values of aTc in competitive and non-competitive circuits
(Fig.~\ref{fig: Variation of mean and FF against GAL}c, and \ref{fig: Variation of mean and FF against GAL}d)
as well, except for aTc = 20 in non-competitive circuit. 
\item Only for the competitive model $FF_{m}^{WTR}>FF_{m}^{WR}$ while\,
$FF_{p}^{WTR}<FF_{p}^{WR}$ for aTc = 20 (Fig.~\ref{fig: Variation of mean and FF against GAL}c
and \ref{fig: Variation of mean and FF against GAL}g). 
\item For aTc$>45$, $FF_{p}^{WR}$ falls below $FF_{p}^{WTR}$, just after
attending a quick peak value at lower GAL concentrations. 
\item The Fano factor, $FF_{m}(FF_{p})$ at $50<aTc<65$ rises sharply then
gets a certain stability (straight horizontal part) for a very short
range of GAL (0.1 - 0.3 unit) and finally falls with increasing GAL.
This feature is absent in other values of aTc. 
\begin{figure}[H]
\begin{centering}
\subfloat[]{\begin{centering}
\centering{}\includegraphics[width=6cm,totalheight=4cm,height=4cm]{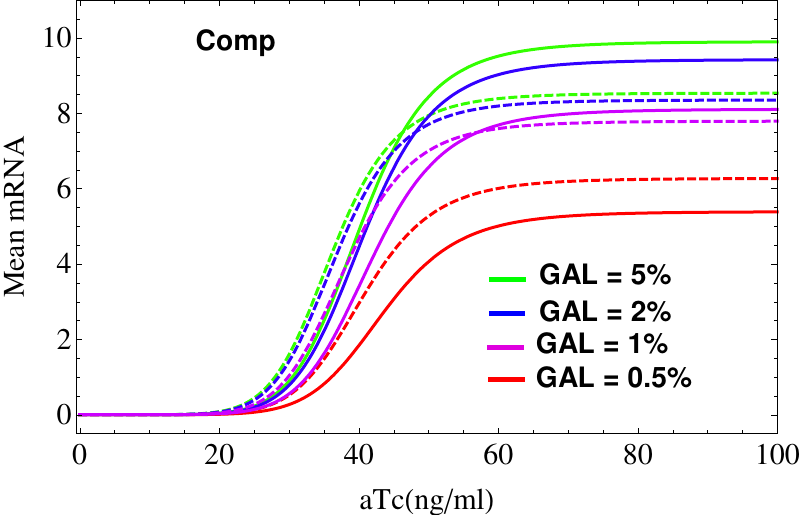}
\par\end{centering}
}\subfloat[]{\begin{centering}
\centering{}\includegraphics[width=6cm,totalheight=4cm,height=4cm]{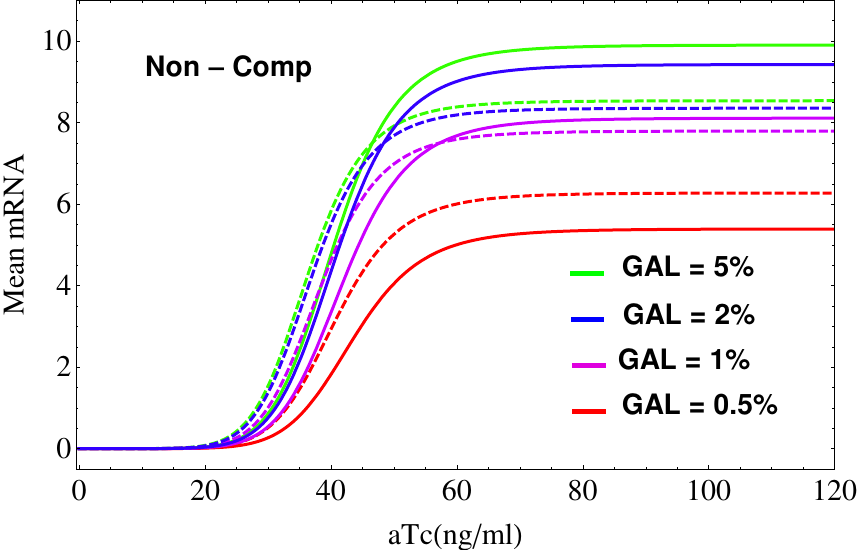}
\par\end{centering}
}
\par\end{centering}
\begin{centering}
\subfloat[]{\begin{centering}
\centering{}\includegraphics[width=6cm,totalheight=4cm,height=4cm]{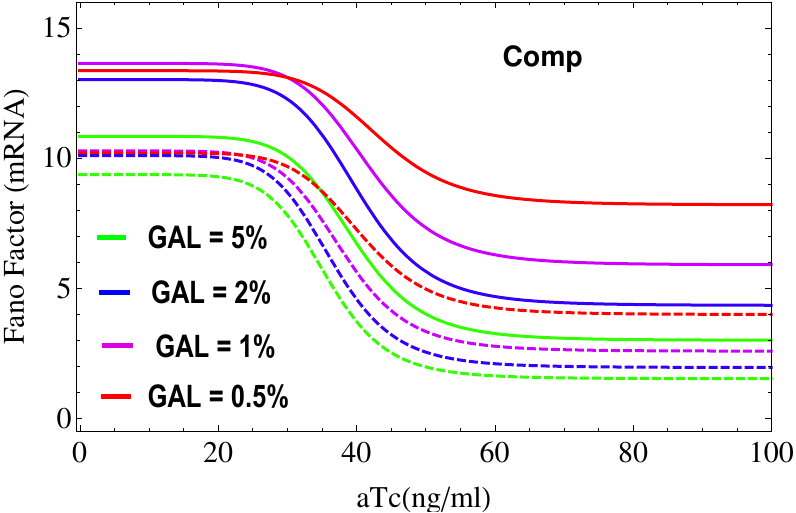}
\par\end{centering}
}\subfloat[]{\begin{centering}
\centering{}\includegraphics[width=6cm,totalheight=4cm,height=4cm]{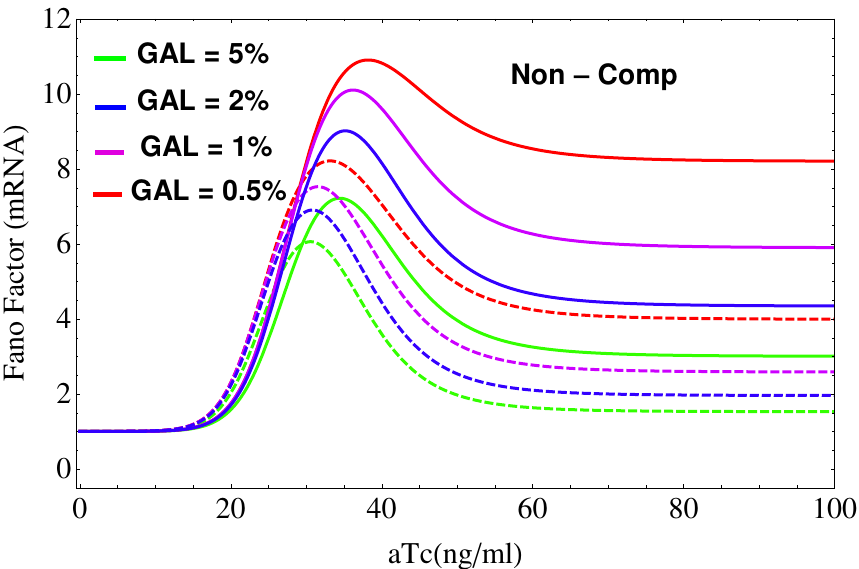}
\par\end{centering}
}
\par\end{centering}
\begin{centering}
\subfloat[]{\begin{centering}
\centering{}\includegraphics[width=6cm,totalheight=4cm,height=4cm]{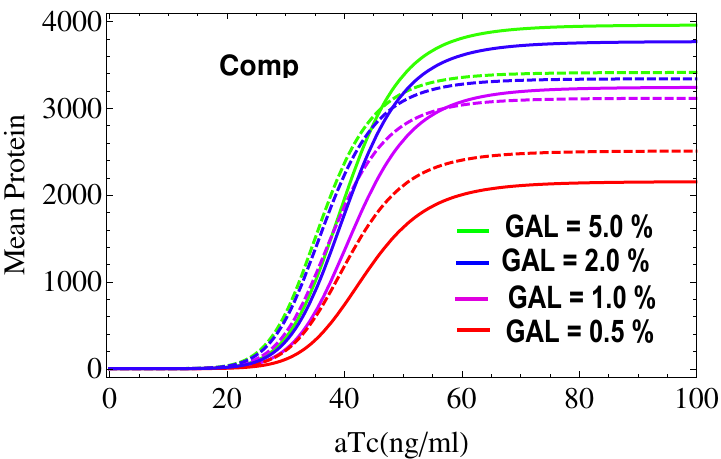}
\par\end{centering}
}\subfloat[]{\begin{centering}
\centering{}\includegraphics[width=6cm,totalheight=4cm,height=4cm]{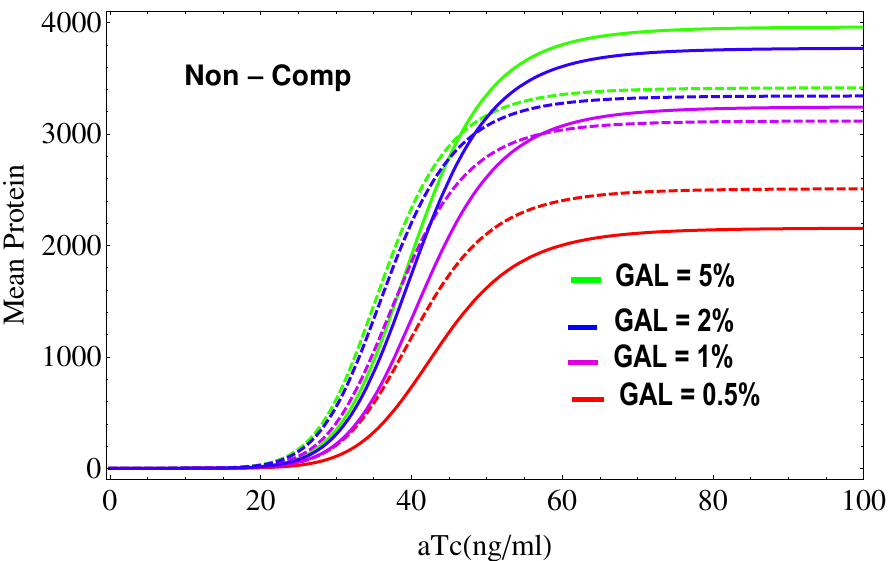}
\par\end{centering}
}
\par\end{centering}
\centering{}\subfloat[]{\begin{centering}
\centering{}\includegraphics[width=6cm,totalheight=4cm,height=4cm]{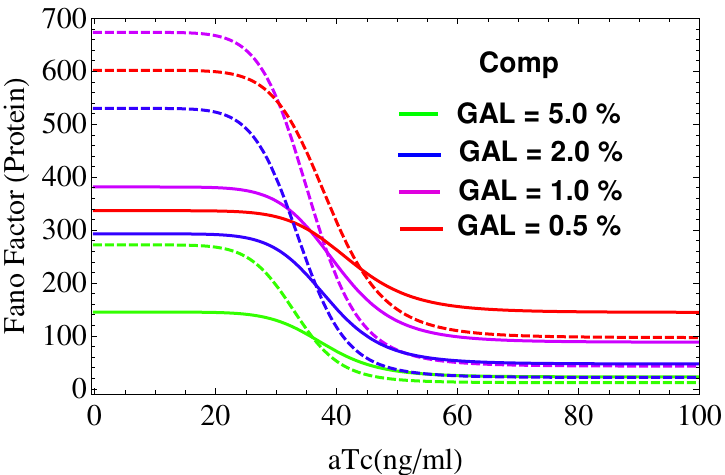}
\par\end{centering}
}\subfloat[]{\begin{centering}
\centering{}\includegraphics[width=6cm,totalheight=4cm,height=4cm]{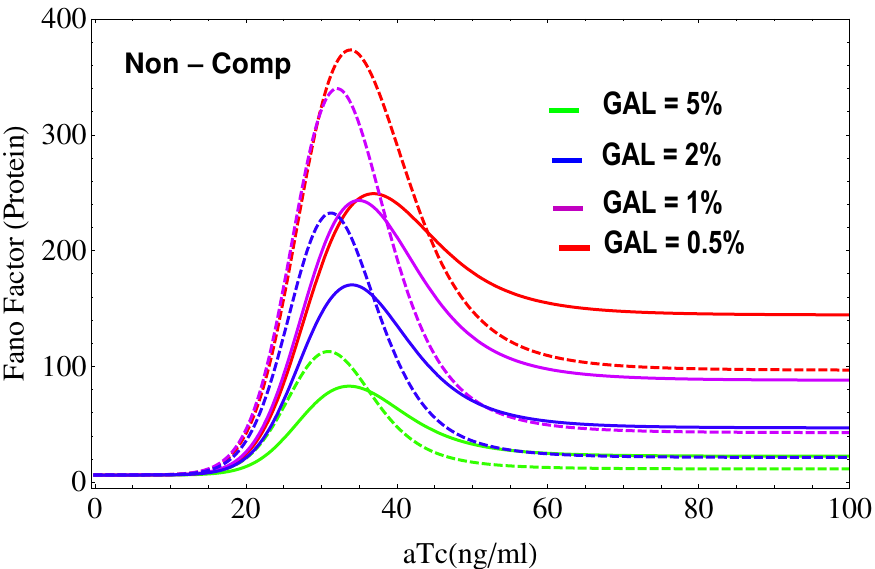}
\par\end{centering}
}\caption{{\small Variation of mean and Fano factor for mRNA and protein level
against aTc with different GAL. The solid (dashed) curves correspond
to the system without (with) reinitiation for reaction scheme \ref{fig:competitive-binding}a
and \ref{fig:Non-competitive-binding}a (\ref{fig:competitive-binding}b
and \ref{fig:Non-competitive-binding}b). The rate constants are chosen
from \cite{blake2003noise} with $J_{m}=15$ and $k_{11}=0$. abbreviation
'Comp (Non-comp)' implies competitive (non-competitive) model.}}\label{fig: Variation of mean and FF against aTc}
\end{figure}
\end{itemize}
The sigmoid nature of the mean expressions with aTc implies that there
exist three regions of operation: cut-off, active or sensitive, and
saturation \cite{das2022stochastic}. A high value of aTc implies
low repression while a lower aTc concentration makes the system OFF
due to high repression. The parameters $k_{2}$ in the competitive
model and $k_{10}$ in the non-competitive model contain \textit{aTc
bounded tetR} act as repressor binding rates. We have found some interesting
features of mean expressions and the Fano factors varying with aTc
for fixed GAL values (colour scheme: GAL = 0.5\%, red; GAL = 1.0\%,
purple; GAL = 2.0\%, blue; GAL = 5.0\%, green) for both the models
(fig. \ref{fig: Variation of mean and FF against aTc}a to \ref{fig: Variation of mean and FF against aTc}h).
Below are some additional observations that we make from our analysis- 
\begin{itemize}
\item At the region of sensitivity ($20\leq aTc\leq60$), $m^{WR}\thinspace(or,\thinspace p^{WR})>m^{WTR}\thinspace(or,\thinspace p^{WTR})$
but $m^{WR}\thinspace(\mbox{or,}\thinspace p^{WR})<m^{WTR}\thinspace(\mbox{or,}\thinspace p^{WTR})$
at saturation region (aTc $>$ 60), only exception holds for curves
with lower GAL concentrations ($\sim0.5\%$). 
\item Fig.~\ref{fig: Variation of mean and FF against aTc}c shows that,
$FF_{m}^{WTR}>FF_{m}^{WR}$ in all three regions, while this behavior
is observed beyond a point of intersection at a certain aTc value
for each pair of curves in the non-competitive model (Fig.~\ref{fig: Variation of mean and FF against aTc}d).
We further notice in Fig.~\ref{fig: Variation of mean and FF against aTc}c
that, the noise (WTR) for GAL$=0.5\%$ (solid red curve) is slightly
lesser than the noise for GAL$=1.0\%$ (solid purple curve ) at the
early stages of aTc concentration ($\sim$ aTc $<35$), after that
it goes higher of GAL =1.0\% curve for higher aTc values. No such
anomaly has been found in the non-competitive system. 
\item It is noteworthy to mention a major difference that exists in the
noise profile for protein level, which arises due to the reinitiation
effect. We notice, $FF_{p}^{WR}>>FF_{p}^{WTR}\,$ at initial values
of aTc and at a particular value of aTc each WR curve crosses over
the WTR curve. Subsequently WR curve goes below the WTR curve at higher
aTc concentrations. A noticeable anomaly is found for Fig.~\ref{fig: Variation of mean and FF against aTc}g
that, curve corresponds to GAL $=0.5\%$ goes slightly lower to the
curve for GAL $=1.0\%$ till aTc $\simeq31.0$ (WTR) and aTc $\simeq$
36.0 (WR) instead of going higher. However, no such anomaly is found
in case of non-competitive circuit. 
\item The amount of noise is higher in competitive circuit than that in
non-competitive one. 
\end{itemize}

\subsection{Noise reducing factors }

\label{subsec:noisered}

\begin{figure}[H]
\begin{centering}
\subfloat[]{\begin{centering}
\centering{}\includegraphics[width=6cm,totalheight=4cm,height=4cm]{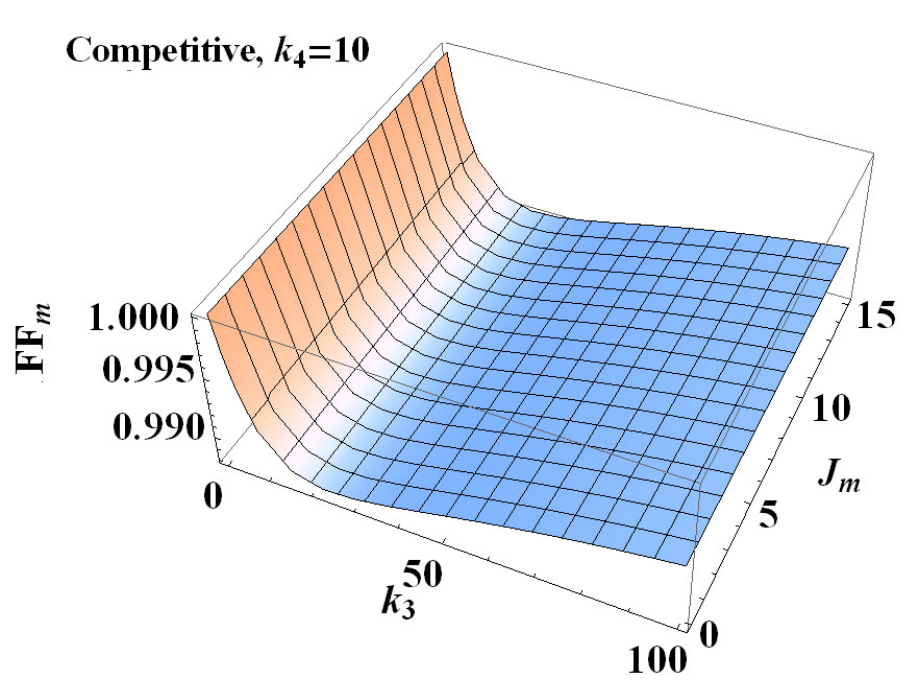}
\par\end{centering}
}\subfloat[]{\begin{centering}
\centering{}\includegraphics[width=6cm,totalheight=4cm,height=4cm]{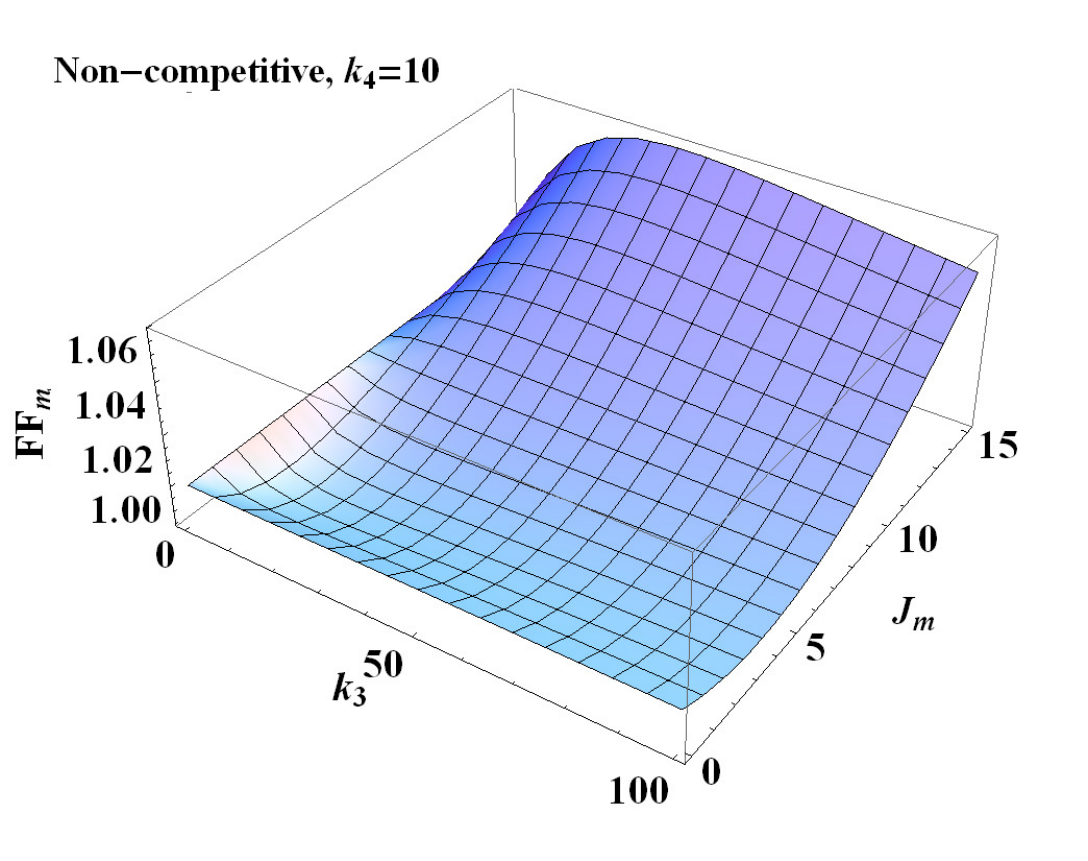}
\par\end{centering}
}
\par\end{centering}
\centering{}\subfloat[]{\begin{centering}
\centering{}\includegraphics[width=6cm,totalheight=4cm,height=4cm]{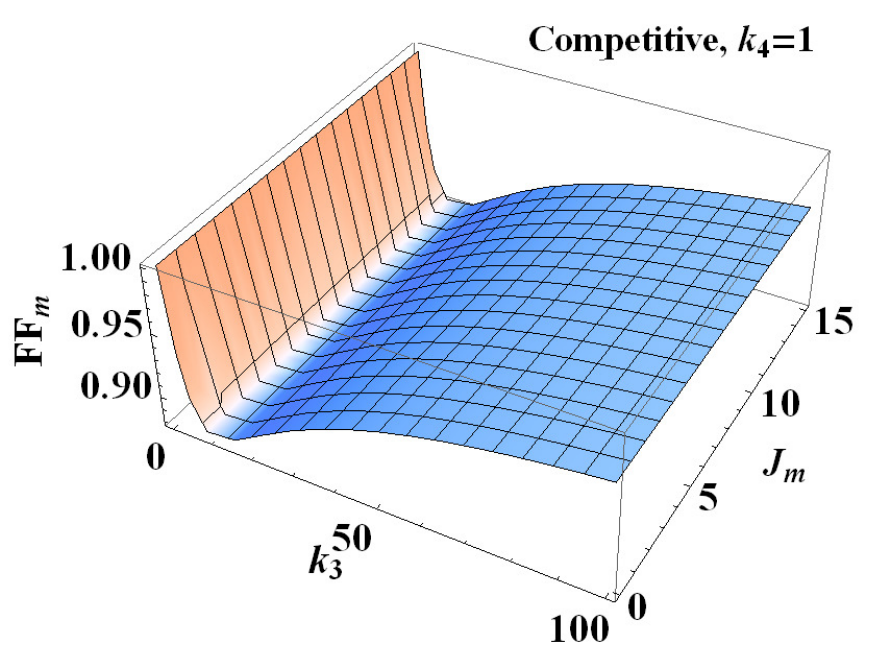}
\par\end{centering}
}\subfloat[]{\begin{centering}
\centering{}\includegraphics[width=6cm,totalheight=4cm,height=4cm]{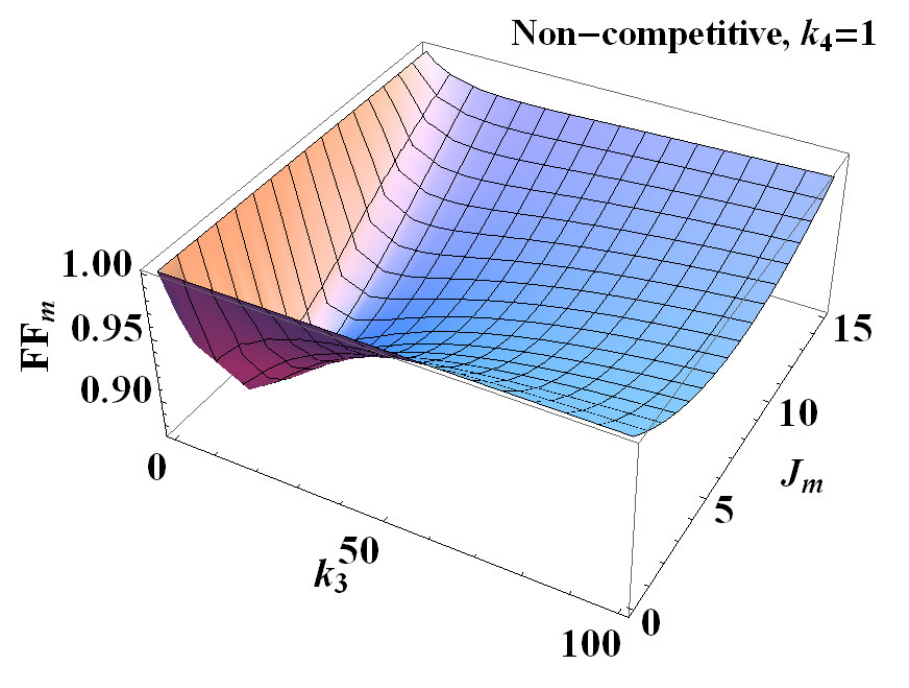}
\par\end{centering}
}\caption{{\small Role of $k_{4}$ in reducing noise strength: Variation of the
Fano factor at mRNA level with $k_{3}$ and $J_{m}$ (a) competitive
circuit with $k_{4}=10$, (b) non-competitive circuit with $k_{4}=10$,
(c) competitive circuit with $k_{4}=1$ and (d) non-competitive circuit
with $k_{4}=1.$ Other parameter values are same as in \cite{blake2003noise}
with GAL = 30\%, aTc = 65 ng/ml, $J_{m}=15$ and $k_{11}=0.$ These
figures are obtained when reinitiation is present in the system.}}
\label{fig:Role-of-k4} 
\end{figure}

We notice that noise can be enhanced or reduced by reinitiatian from
Fig.~\ref{fig: Variation of mean and FF against GAL} and Fig.~\ref{fig: Variation of mean and FF against aTc}.
It is also visible that the competitive binding model shows more noise
than the non-competitive one. We now attempt to reduce the noise much
below the Poissonian level ($FF_{m}=1)$. We see, Fig. \ref{fig:Role-of-k4}
reveals that it is possible if the value of $k_{4}$ is reduced from
$10$ to $1$. According to Blake et al. \cite{blake2003noise} when
$k_{4}=1$ the RNAP-II binds much more tightly with the promoter even
when the repressor is rarely present. This results in much reduction
of noise for the circuits of both types. Moreover, we find that noise
in the competitive model can be reduced more than non-competitive
one in the sub-Poissonian region ($FF_{m}<1)$, corresponding to $k_{4}=10$
(Figs.~\ref{fig:Role-of-k4}a and \ref{fig:Role-of-k4}b) or $k_{4}=1$
(Figs.~\ref{fig:Role-of-k4}c and \ref{fig:Role-of-k4}d).

\section{Conclusions and Discussions}

\label{sec:conc} The regulation of gene expression through transcription
factors (TFs) and control of noise are biologically significant \cite{magee2003haploinsufficiency,weinberger2005stochastic,dar2014screening}
to respond to the intra-cellular and external environmental changes
as well \cite{alberts2002helper}. They are also important in various
medical applications \cite{magee2003haploinsufficiency,weinberger2005stochastic,dar2014screening}.
For example, the TFs (activators and repressors) can act as a tumor
suppressor in prostate cancer \cite{magee2003haploinsufficiency}.
The noise, arising due to various stochastic events in gene expression
is used to modulate the reactivation of HIV from latency (a quiescent
state, that is a major barrier to an HIV cure). Noise suppressors
stabilize latency while noise enhancers reactivate latent cells \cite{dar2014screening}.
On the contrary, the noisy expression can be beneficial to a population
of genetically identical cells by composing phenotypic heterogeneity
\cite{blake2006phenotypic,thattai2004stochastic,kaern2005stochasticity,raj2008nature,alberts2002helper}.

In this paper, we have studied a gene transcription regulatory architecture
where activator and repressor molecules bind the promoter of gene
in a competitive fashion. There may be various mode of competition
among the TFs and promoter binding sites \cite{karmakar2010conversion,das2017effect}.
Some theoretical studies have explored the kinetics and recorded the
effects of TFs on noise \cite{das2017effect,burger2012influence,soltani2015nonspecific}.
Meanwhile, we found an experimental study on a competitive binding
network performed by Rossi et al. \cite{rossi2000transcriptional}.
Almost a decade later, Yang and Ko \cite{yang2012stochastic} tried
to explain the experimental results with the help of stochastic simulation.
However, both approaches were unable to explain the collective (Hill
coefficient) and stochastic behavior of the three-state competitive
activator-repressor system because of the lack of detailed kinetics
and parameter values. A pioneering work on the same model was done
by Karmakar \cite{karmakar2010conversion}, who also found the similar
distributions (PDF) as in \cite{rossi2000transcriptional} by theoretical
analyses and explained how the response changes from graded to binary
in presence of either activator or repressor or both. However, the
required parameter values or the exact relationship of the parameters
with dox (stimulus) were still unavailable. Moreover, the above-mentioned
works did not focus on the noise profile of the three-state competitive
activator-repressor system. In recent work, Braichencho et al. \cite{braichenko2021distinguishing}
modeled a network where they have studied a proximal promoter-pausing
in a three-state activator-repressor system, which can be effectively
described by a two-state model in some limiting case. Another competitive
model was proposed in \cite{jiao2020regulation} where the regulation
of gene expression can occur via two cross talking parallel pathways.
Although, our approaches and investigations are different in the present
work.

The goal of our study is to fabricate a framework to explain the model
with a proper theory supported by analytical and/or simulation methods.
Consequently, we developed a theory that offers detailed chemical
kinetics of the competitive activator-repressor system and a most
probable set of parameter values that fit the experimental data \cite{rossi2000transcriptional}
quite nicely (Fig.~\ref{fig:AR Curve-fitting_all}b). The dox-dependent
parameters, here we found analytically, having very much similar form
that proposed earlier in \cite{blake2003noise} instead of the Hill
function\cite{rossi2000transcriptional,yang2012stochastic}. However,
in \cite{blake2003noise} the authors did not clearly mention how
the structures of these parameters are formed. Hence, we have designed
an analytical treatment by assuming intermediate gene-dox complex
formation to determine the exact forms of the parameters (\textit{i.e.}
relation with dox) and revealed how the mean expression and noise
vary with the changes of these parameter values. The robustness of
our parameter estimation is supported by the minimization of relative
error and the mean squared error given in Appendix-B. We have found
the important parameters that govern the stochastic noise of the circuit.
On top of that, we have made a comparison of characteristics between
competitive and non-competitive architecture. We have also noticed
some anomalous behavior of both mean and the Fano factor with the
variation of activator (GAL) and/or repressor (aTc) whenever reinitiation
of transcription was taken under consideration. We have noticed that,
just as the noise in competitive circuit is high in the super-Poissonian
region, its noise can be reduced much lower in the sub-Poissonian
regime in comparison with the the non-competitive one. However, the
mean expression levels remain the same under identical set of rate
constants (refer to Fig.\,\ref{fig: Variation of mean and FF against GAL},
\ref{fig: Variation of mean and FF against aTc}, and \ref{fig:Role-of-k4}).
The reason behind such behaviour is subject to further analysis which
we have not performed here.

Our proposed theory may be applicable in synthetic biology to understand
the architecture of interactions which may buffer the stochasticity
inherent to gene transcription of a complex biological system. By
using our analytical approach, it is possible to predict the mean
and standard deviation of the number of transcribed mRNAs/proteins.
There can be further extensive study on the model considering the
effect of other parameters that are not considered in present analysis.
The availability of a well-designed analytical theory with detailed
reaction kinetics of that model could be the powerful tools for future
research and analysis for the similar or comparable genetic networks.

\subsection*{Statements and Declarations}

Competiting Interests: The authors declare that they have no known
competing financial interests or personal relationships that could
have appeared to influence the work reported in this paper.

\subsection*{Acknowledgment }

The authors would like to acknowledge their colleagues and friends
for their consistent support and encouragement to this research work.

\subsection*{Glossary}
\begin{itemize}
\item \textbf{Transcription factor: }Transcription factors (TFs) are proteins
that have DNA binding domains with the ability to bind to the specific
sequences of DNA (called promoter). They control the rate of transcription.
If they enhance transcription they are called activators and repressors
if they inhibit transcription. 
\item \textbf{Fano factor and Noise strength: }The Fano factor is the measure
of deviations of noise from the Poissonian behavior and is defined
as \cite{das2022stochastic,braichenko2021distinguishing}, 
\[
Fano\,factor\,(mRNA)=\frac{variance\thinspace of\,protein}{mean\,mRNA}=\frac{(standard\,deviation)^{2}}{mean}.
\]
So, for a given mean, smaller the Fano factor implies smaller variance
and thus less noise. Therefore, the Fano factor gives a measure of
noise strength which is defined (mathematically) as \cite{blake2003noise},
\[
noise\ strength=\frac{variance\thinspace of\,protein}{mean\,mRNA}=\frac{(standard\,deviation)^{2}}{mean}.
\]
So, the Fano factor and noise strength are synonymous. 
\item \textbf{Transcriptional efficiency : }Transcriptional efficiency is
the ratio of instantaneous transcription to the maximum transcription
\cite{blake2003noise}. 
\end{itemize}

\subsection*{Appendix A}

\label{part:Appendix-moment generating fun}

In an attempt to find out the expressions of mean mRNA (protein) and
the corresponding Fano factors, we have used a moment generating function
which is defined as 
\begin{equation}
F(z_{i},t)=\sum_{n=0}^{\infty}z_{i}^{n_{i}}P(n_{i},t).\label{eq:A1_generating fun}
\end{equation}
Here, $i=1,2,......,5$.

We have, 
\begin{align}
\frac{\partial F(z_{i},t)}{\partial t} & =\sum_{n=0}^{\infty}z_{i}^{n_{i}}\frac{\partial P(n_{i},t)}{\partial t},\nonumber \\
 & =k_{1}(z_{1}-1)\left[lF-z_{1}\frac{\partial F}{\partial z_{1}}-z_{2}\frac{\partial F}{\partial z_{2}}-z_{3}\frac{\partial F}{\partial z_{3}}\right]\nonumber \\
 & +k_{2}(1-z_{1})\frac{\partial F}{\partial z_{1}}+k_{a}(z_{2}-z_{1})\frac{\partial F}{\partial z_{1}}+k_{d}(z_{1}-z_{2})\frac{\partial F}{\partial z_{2}}\nonumber \\
 & +k_{3}(z_{3}-z_{2})\frac{\partial F}{\partial z_{2}}+k_{4}(z_{2}-z_{3})\frac{\partial F}{\partial z_{3}}\label{eq:A.2_generating fun moments}\\
 & +J_{m}(z_{2}z_{4}-z_{3})\frac{\partial F}{\partial z_{3}}+k_{m}(1-z_{4})\frac{\partial F}{\partial z_{4}}\nonumber \\
 & +J_{p}(z_{5}-1)z_{4}\frac{\partial F}{\partial z_{4}}+k_{p}(1-z_{5})\frac{\partial F}{\partial z_{5}}.\nonumber 
\end{align}

In steady state, $\frac{\partial F(z_{i},t)}{\partial t}=0$ and for
total probability, $F(z_{i}=1,0)=1$.

Now, by setting $\left[\frac{\partial}{\partial z_{1}}(\frac{\partial F}{\partial t})\right]_{z_{i}=1}=0$,
we get $\frac{\partial F}{\partial z_{1}}=f_{1}(say)=\thinspace<n_{1}>\thinspace=$
average number of gene at state $G_{n}$.

similarly, by setting {[}$\frac{\partial}{\partial z_{1}}(\frac{\partial^{2}F}{\partial z_{1}\partial t})]_{z_{i}=1}=0$
will give $\frac{\partial^{2}F}{\partial z_{1}^{2}}=f_{11}(say)$
and so on. Proceeding in the same way we obtain,

$f_{5}=<n_{4}>=mean\thinspace\thinspace mRNA$ and $f_{5}=\thinspace<n_{5}>\thinspace=mean\thinspace Protein$

\[
Fano\thinspace factor\thinspace(mRNA)=\frac{variance\thinspace of\thinspace mRNA}{mean\thinspace mRNA}=\frac{f_{44}+f_{4}-f_{4}^{2}}{f_{4}},
\]

\[
Fano\thinspace factor\thinspace(Protein)=\frac{variance\thinspace of\thinspace Protein}{mean\thinspace Protein}=\,\frac{f_{55}+f_{5}-f_{5}^{2}}{f_{5}},
\]

In addition, if the contributions of basal rate $J_{0}$ and the reverse
transition $k_{R}$ (or $k_{11})$ are taken under consideration,
there would be two additional terms in Eq. (\ref{eq:A.2_generating fun moments})
as,

\[
J_{0}\left(z_{4}-z_{1}\right)\frac{\partial F}{\partial z_{1}},~~~~\mbox{and}~~~~k_{R}\left(z_{1}-z_{3}\right)\frac{\partial F}{\partial z_{3}}.
\]

\subsection*{Appendix B}

\label{fitting parameters and uncertainties} 
\begin{itemize}
\item \textbf{Model fitting parameters:} We proposed an analytical theory
that fits very much to the experimental result, performed in \cite{rossi2000transcriptional}.
The dox-dependent parameters, here we found, having a very similar
form, proposed by Blake et al. in \cite{blake2003noise} instead of
Hill type \cite{rossi2000transcriptional,yang2012stochastic}. However,
in \cite{blake2003noise} the authors did not clearly mention how
the structures of these parameters are formed. Hence, we design an
analytical treatment as described in the main text to determine the
forms of the parameters $k_{a}$, $k_{1}$, $k_{2}$ and $k_{d}$.
While for the best estimates of other parameters are revised through
the minimization of the sum of squared error (SSE) and the mean square
error (MSE). We also provide the parameter sensitivity plot in Fig.
\ref{Fig.- sensitivity of parameters}a. Moreover, we calculate and
check the relative percentile error (RE) for each curve by using the
formula, 
\begin{equation}
RE=\frac{y_{T}-y_{E}}{y_{E}}100\%,
\end{equation}
where, \textit{T} stands for theoretical and \textit{E} stands for
experimental. 
\item \textbf{Uncertainty in fitted parameters :} The uncertainty of the
parameter estimation, is generally expressed by the mean square errors,
is proportional to the SSE (SSWE) and inversely proportional to the
square of the coefficient of sensitivity of the model parameters \cite{smith1997numerical}.
The mean square fitting error is 
\begin{equation}
\sigma^{2}=\frac{1}{n-r}\sum_{i=1}^{n}\left[w_{i}(y_{T}-y_{E})\right]^{2}=\frac{Sum\thinspace of\thinspace squared\thinspace weighted\thinspace error}{(n-r)},
\end{equation}
where $n$ is the number of observations and \textit{r} is the number
if parameters are being determined. The weighting factor, $w_{i}$
is determined by the slope of the curve at each data point. 
\end{itemize}
The sensitivity ($\mathcal{S}$) of a function $f(r)$ over the parameter
$r$ is given by 
\begin{equation}
\mathcal{S}=\frac{r}{f(r)}.\frac{\partial f(r)}{\partial r}.
\end{equation}

We obtain the sensitivity of the Fano factor (mRNA) over the fitting
parameters and calculate MSE of each parameter keeping others as constant.
\begin{equation}
MSE=\frac{\sigma^{2}}{\sum_{i=1}^{n}\left[\frac{\partial f(r)}{\partial r}\right]^{2}},
\end{equation}
where the denominator is the coefficient of sensitivity, squared and
summed over all observations. 
\begin{center}
\begin{figure}[H]
\begin{centering}
\subfloat[]{\begin{centering}
\centering{}\includegraphics[width=12cm,totalheight=8cm]{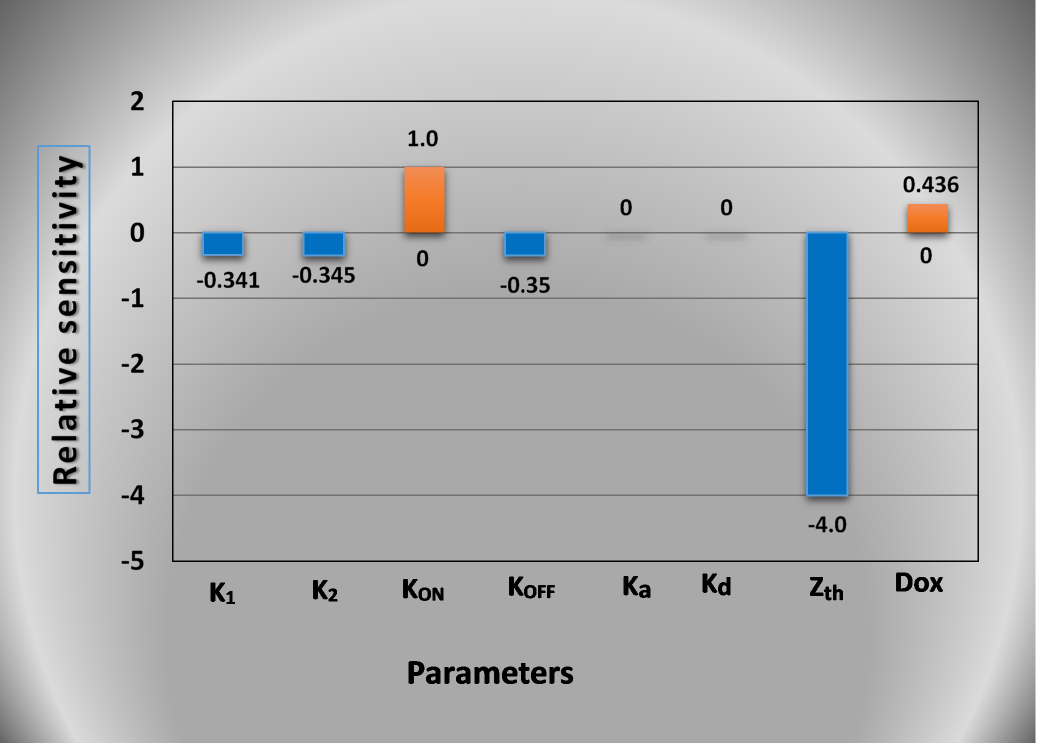}
\par\end{centering}
}
\par\end{centering}
\centering{}\subfloat[]{\begin{centering}
\includegraphics[width=12cm,totalheight=8cm]{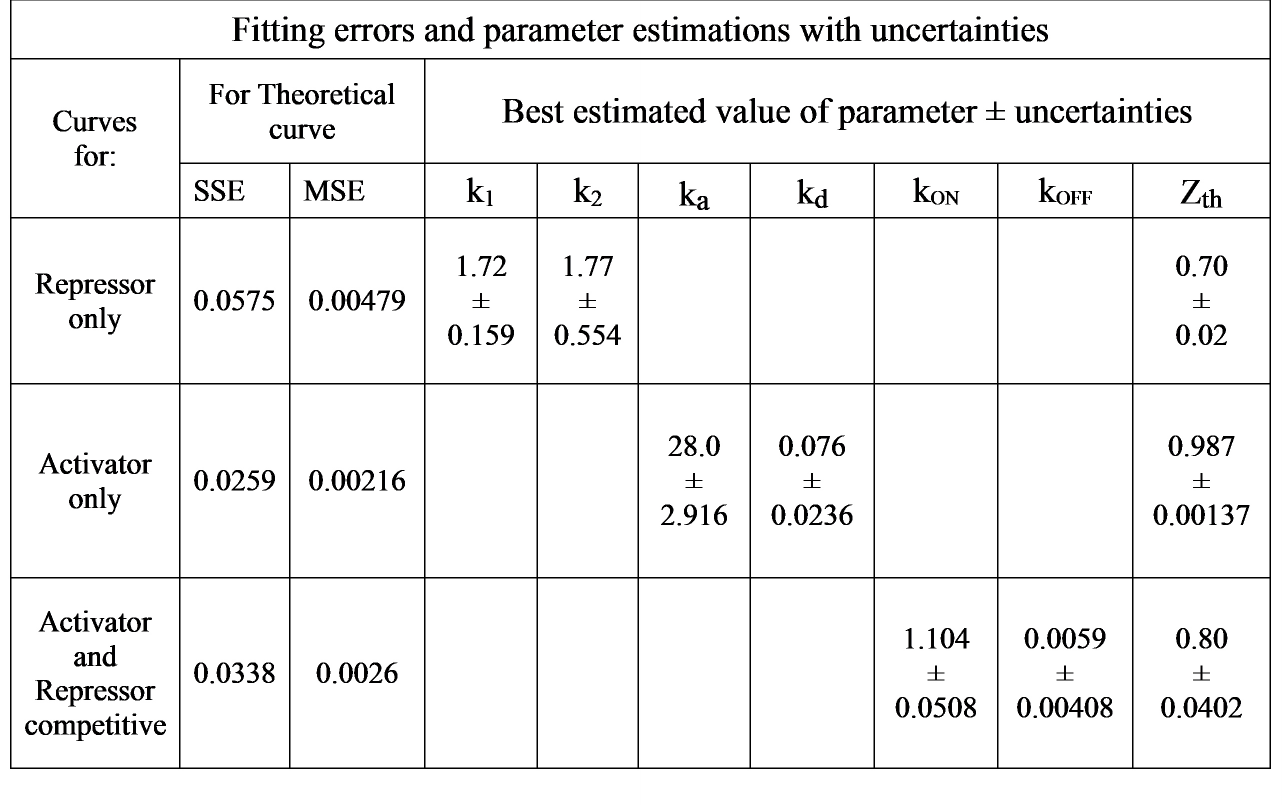} 
\par\end{centering}
}{\small\caption{{\small (a) Sensitivity of parameters, (b) Minimized error of fitting
and parameter estimation with uncertainties.~}}
\label{Fig.- sensitivity of parameters}}{\small\par}

\end{figure}
\par\end{center}

\begin{center}
We also notice that sensitivity is independent of $J_{1}$, $J_{0}$,
$k_{a}$ and $k_{d}$ as well. The $Z_{th}$ is the most sensitive
parameter and it was kept constant for each case. We found that the
promoter activity is sensitive within a small range of values of the
chosen parameters. The best parameter estimation and minimization
of errors (see Fig.~\ref{Fig.- sensitivity of parameters}b) support
the robustness of our result. The square root of the MSE is the standard
deviation, and the approximate 95\% confidence interval for \textit{r}
is \cite{smith1997numerical} 
\begin{equation}
[r]_{95\%}=\mathcal{R}\pm2\surd MSE,
\end{equation}
$\mathcal{R}$ is the best estimate value of parameter $r.$ 
\par\end{center}

\end{document}